# APPROXIMATING (UNWEIGHTED) TREE AUGMENTATION VIA LIFT-AND-PROJECT, PART I: STEMLESS TAP

JOSEPH CHERIYAN AND ZHIHAN GAO

Abstract. In Part I, we study a special case of the unweighted *Tree Augmentation Problem* (TAP) via the Lasserre (Sum of Squares) system. In the special case, we forbid so-called stems; these are a particular type of subtree configuration. For stemless TAP, we prove that the integrality ratio of an SDP relaxation (the Lasserre tightening of an LP relaxation) is $\leq \frac{3}{2} + \epsilon$, where $\epsilon > 0$ can be any small constant. We obtain this result by designing a polynomial-time algorithm for stemless TAP that achieves an approximation guarantee of $(\frac{3}{2} + \epsilon)$ relative to the SDP relaxation. The algorithm is combinatorial and does not solve the SDP relaxation, but our analysis relies on the SDP relaxation.

We generalize the combinatorial analysis of integral solutions from the previous literature to fractional solutions by identifying some properties of fractional solutions of the Lasserre system via the decomposition result of Karlin, Mathieu and Nguyen (IPCO 2011).

Also, we present an example of stemless TAP such that the approximation guarantee of $\frac{3}{2}$ is tight for the algorithm.

In Part II of this paper, we extend the methods of Part I to prove the same results relative to the same SDP relaxation for TAP.

## 1. Introduction

In the weighted *Tree Augmentation* problem we are given a connected, undirected multigraph $G$ with non-negative costs on the edges, together with a spanning tree $T$ of $G$. The goal is to find a set of edges, $F \subseteq E(G) - E(T)$, of minimum cost such that the multigraph $(V, E(T) \cup F)$ is 2-edge-connected. By a link we mean an element of $E(G) - E(T)$; thus, a link is an edge of $G$ that can be used to augment $T$. We say that a link $uw$ covers an edge $\hat{e} \in E(T)$ (a tree-edge) if the multigraph $T + uw - \hat{e} = (V, E(T) \cup \{uw\} - \{\hat{e}\})$ is connected. * We say that a set of links $F$ covers the tree $T$ if every edge of $T$ is covered by at least one link of $F$; it can be seen that $F$ covers $T$ iff the multigraph $(V, E(T) \cup F)$ is 2-edge-connected. Thus, the goal is to compute a set of links of minimum cost that covers $T$.

The weighted Tree Augmentation problem was first studied by Frederickson and Jaja in 1981 [7]. They showed that the problem is NP-hard, and they presented a 2-approximation algorithm. Subsequently, it has been proved that even the unweighted Tree Augmentation problem is APX-hard, see [13, Section 4]; thus the unweighted problem has no PTAS assuming P≠NP.

There have been some important advances on this problem for the corresponding *unweighted* (i.e., uniform weight) problem. Following [9], we use the abbreviation TAP for the *unweighted* Tree Augmentation problem. Nagamochi, see [10], presented the first algorithm for TAP that improved on the approximation guarantee of 2; the approximation guarantee is $\approx 1.875$. Subsequently, Even, et al., [9] built on the ideas and techniques initiated by Nagamochi and presented an elegant algorithm and analysis that achieves an approximation guarantee of 1.8. Even, et al., explain that the threshold of 1.8 is a natural barrier for TAP; this comes from a particular type of subtree configuration, a so-called "stem" , that occurs in instances of TAP (see Section 2 for definitions). To improve on the approximation guarantee of 1.8, both the algorithm and analysis have to be refined to handle stems. This introduces significant new complications. In a conference publication

*In other words, a link covers every tree-edge in the unique path of $T$ between the ends of the link.

from 2001, Even, et al., reported a 1.5 approximation algorithm for TAP, see the extended abstract [5], and recently, Kortsarz & Nutov [15] presented the journal version of this result.

There are several other papers on TAP, see e.g., [1, 17, 3, 6, 8], but we do not discuss them since either they are not directly relevant to this paper or they are unpublished manuscripts. †

Linear programming relaxations for the weighted version of TAP have been studied for many years. (We use the standard abbreviation LP to mean a linear programming relaxation or a linear programming problem.) There is an obvious "covering" LP: we have a variable $x_e$ for each edge $e \in E(G) - E(T)$ and we have a covering constraint for each edge of $T$. It is well known that the integrality ratio of this LP is $\leq 2$; this can be deduced from Jain's result [11]. A lower bound of 1.5 on the integrality ratio is known [2]; in fact, the construction for the lower bound uses uniform weights for the edges in $E(G) - E(T)$, hence, the lower bound applies for TAP.

**Our results and techniques.** In Part I, our main contribution is to apply the Lasserre system to a special case of TAP, and to derive some properties of the feasible solutions that are then used to analyze the integrality ratio and the approximation guarantee. A stem is a node incident to three edges of $T$ that satisfies some additional conditions (see Figure 2 for an illustration, and see Section 2 for formal definitions). We use *stemless TAP* to refer to the special case of TAP where the instance is such that the tree $T$ has no stem. In Part I, we focus on stemless TAP, and prove that the integrality ratio of an SDP relaxation (the Lasserre tightening of an LP relaxation) is $\leq \frac{3}{2} + \epsilon$, where $\epsilon > 0$ can be any small constant.

The Lasserre system applies to an initial LP, and it derives a sequence of tightenings of the initial LP; these tightened relaxations are indexed by a number $t = 0, 1, \ldots$ called the *level*, where the level 0 tightening means the initial LP. A key "decomposition theorem" (see Theorem 4.1, [18, 12]) asserts that a feasible solution at level $t$ can be written as a convex combination of feasible solutions at a lower level such that all of these lower-level solutions $y$ are "locally integral." Here, "locally integral" means that there is a specified subset $J \subseteq E(G) - E(T)$, such that the solution $y$ takes only zero or one values on this subset (i.e., $y_e \in \{0, 1\}$, $\forall e \in J$). A key point is that the difference in levels (between the level $t$ of the given feasible solution and the level of the lower-level "locally integral" solutions) does *not* depend on the size of $J$, rather, it depends on the following "combinatorial parameter" determined by $J$. Suppose that there exists a constant $k$ such that every feasible solution $x$ of the initial LP has $\leq k$ entries in $J$ that have value one, i.e., $|\{e \in J \,:\, x_e = 1\}| \leq k$ for every feasible solution $x$ of the initial LP. Then for any $t > k$, a feasible solution at level $t$ can be written as a convex combination of "locally integral" feasible solutions at level $(t - k)$. This property does not hold for other weaker Lift-and-Project systems such as the Lovász-Schrijver system or the Sherali-Adams system.

We formulate an initial LP that is a tightening of the obvious "covering" LP for TAP; see $(LP_0)$ in Section 3. For this purpose, we introduce the notion of overlapping pairs, and we add a family of constraints on overlapping pairs to our initial LP; see Proposition 3.1 in Section 3. This (together with the decomposition theorem) turns out to be the key for proving properties of feasible solutions to the Lasserre system.

The analysis of our algorithm is based on a potential function. Our potential function is derived from the Lasserre tightening of the initial LP. In contrast, the previous literature uses potential functions that are derived from combinatorial lower bounds. We present an example showing that our potential function is not valid for our initial LP; in other words, the Lasserre tightening is essential for one part of our analysis.

---

†A preliminary version of our paper with a weaker approximation guarantee of $(1.75 + \epsilon)$ for TAP relative to the same SDP relaxation was circulated widely in July 2014, and it led to subsequent publications by others, e.g., [14], but we prefer to avoid discussion on these matters and we leave it to the subsequent publications.

Our algorithm is "combinatorial" and we do *not* need to solve the initial LP nor its Lasserre tightening to run the algorithm (but, the analysis of the algorithm relies on the Lasserre tightening). Our algorithm may be viewed as a variant of the algorithm of [9, Section 3.4]; see Section 6 for details. The algorithm is a greedy-type iterative algorithm that makes a leaves-to-root scan over the tree $T$ and (incrementally) constructs a set of links $F$ that covers $T$. The algorithm starts with $F := \emptyset$, at each major step it adds one or more links to $F$ (it never removes links from $F$), and at termination, it outputs a set of links $F$ that covers $T$ such that $|F| \leq$ the potential function. The algorithm incurs a cost of one unit for each link added to $F$. The key to the analysis is to show that for each major step, the cost incurred (i.e., one plus the number of links added to $F$) is compensated by a part of the potential function; see Section 6 for details.

It is possible that the naive algorithm gets "stuck." But, in this scenario, we can prove that there exists a small combinatorial obstruction. The algorithm can be modified for this scenario. The modified algorithm finds each occurrence of the small combinatorial obstruction in polynomial time, and then handles all of these occurrences in an appropriate way; see Section 7.2 for details; this part of our paper is similar to [9, Section 4.3].

Informally speaking, our analysis in Section 7 asserts the following:

> if the naive algorithm gets "stuck" then there exists a small combinatorial obstruction, a so-called deficient 3-leaf tree, see Theorems 7.3, 7.4.

This assertion is the key to this paper; it turns out that the algorithmic aspects as well as the analysis of the approximation guarantee are straightforward consequences. Our analysis in Section 7 makes essential use of the Lasserre system and the decomposition theorem; see Figure 1 and also see Section 10.

Finally, we present an example of stemless TAP such that the approximation guarantee of $\frac{3}{2}$ is tight for the algorithm.

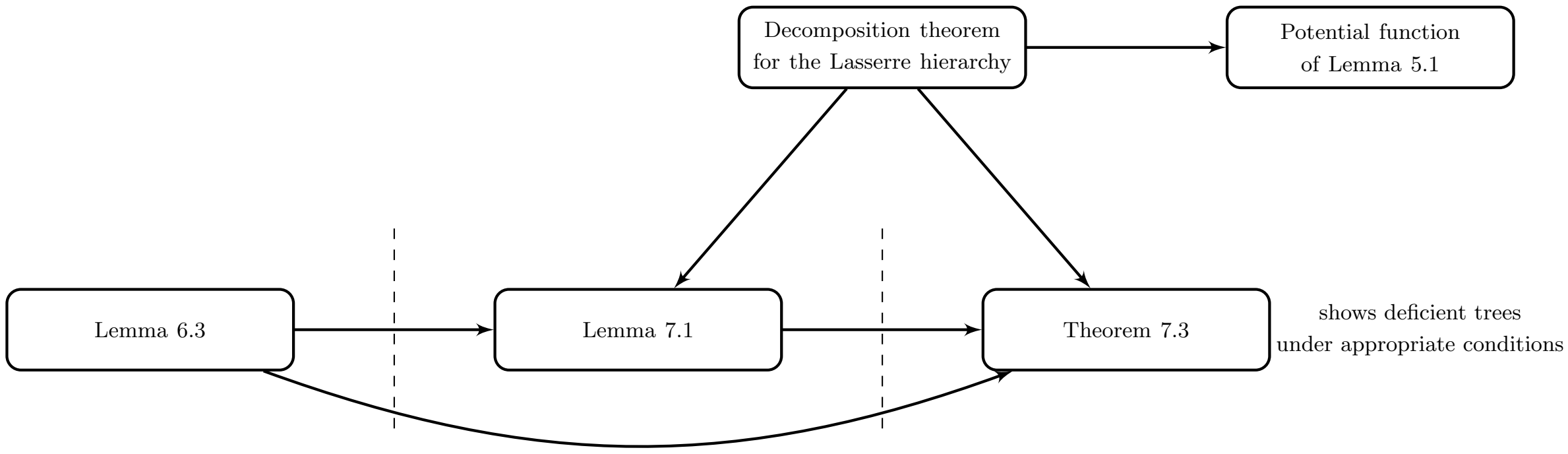

FIGURE 1. An illustration of the role of the decomposition theorem for the Lasserre system (Theorem 4.1) in our analysis. The analysis for Theorem 7.3 consists of three blocks: low-level assertions maintained by the algorithm, intermediate-level results (on credits and structural properties), and the high-level analysis (proof of Theorem 7.3); the three blocks are shown from left to right in the figure. The decomposition theorem pertains to the two right-most blocks. Also, we use the decomposition theorem to derive our potential function (Lemma 5.1).

In Part II of this paper, we extend the methods of Part I to prove the same approximation guarantee for (general) TAP. Part II follows the same outline as Part I. Moreover, the initial LP and Lasserre tightening are the same for both parts. We mention that Part II is essentially self-contained and is "logically independent" of Part I; our motivation for writing Part I is to give an accessible presentation of the algorithmic ideas and correctness arguments used in Part II.

An outline of Part I is as follows. Section 2 has definitions and notation. We adopt the notation and terms of Even, et al., [9], where possible; this will aid readers familiar with that paper. Section 3 presents the initial LP, while Section 4 discusses the Lasserre tightening of the initial LP and proves some basic properties and inequalities. Section 5 derives our potential function, based on a solution $y$ of the Lasserre tightening; this section also has an example showing that our potential function is not valid for the initial LP. Section 6 presents the algorithm and the "credit scheme" used by the algorithm. The most important component of Part I is Section 7; this section proves the key theorem on deficient trees (Theorem 7.3); this section also presents and proves the last piece of the algorithm, namely, the handling of deficient trees. Section 8 presents an example of stemless TAP such that the approximation guarantee of $\frac{3}{2}$ is tight for the algorithm. The last section, Section 10, has conclusions.

## 2. Preliminaries and notation

This section presents definitions and notation.

**Standard notation including tree $T$, link set $E$.** Let $G = (V, E(G))$ be a connected, undirected multigraph, and let $T = (V, \widehat{E}_T)$ be a spanning tree of $G$. We assume that $|V| \geq 2$. By a *tree-edge* we mean an edge of $T$. Let $E$ denote the edge-set $E(G) - \widehat{E}_T$; we call $E$ the *link set* and we call an element $\ell \in E$ a *link*; thus, a link is an edge of $G$ that can be used to augment $T$. An instance of TAP consists of $G$ and $T$. We assume that all instances of interest have feasible solutions, that is, we assume that $(V, \widehat{E}_T \cup E)$ is 2-edge-connected. The goal is to find a minimum-size subset $F$ of $E$ such that augmenting $T$ by $F$ results in a 2-edge-connected multigraph, i.e., the multigraph $(V, \widehat{E}_T \cup F)$ is 2-edge-connected.

For two nodes $u, v \in V$, we use $P_{u,v} = P_{v,u}$ to denote the unique path of the tree $T$ between $u$ and $v$.

For a node $v \in V$, we denote the number of tree-edges incident to $v$ by $\deg_T(v)$. Let $uv$ be a link such that $\deg_T(u) = 1$, $\deg_T(v) = 1$, there exists an internal node $s$ in $P_{u,v}$ such that $\deg_T(s) = 3$, and every other node $w$ (if any) in $P_{u,v}$ has $\deg_T(w) = 2$. Then, $uv$ is called a *twin link* and $s$ is called a *stem*.

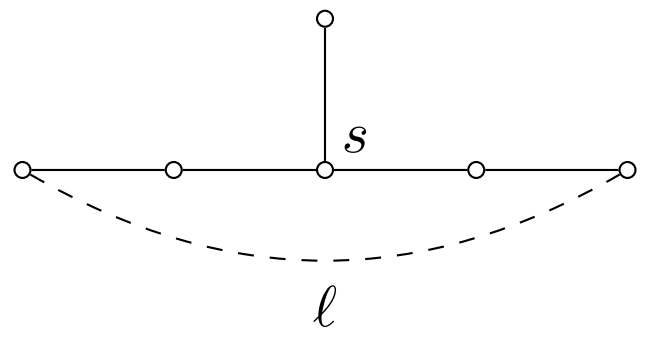

Figure 2. Illustration of a stem $s$ and its twin link $\ell$. The tree-edges are indicated by solid lines and the twin link $\ell$ is indicated by dashed lines.

We use *stemless TAP* to refer to the special case of TAP where the instance is such that the tree $T$ has no stems.

For any $U \subseteq V$, we denote the set of links with both ends in $U$ by $E(U)$, and for any two subsets $U, W$ of $V$, we denote the set of links with one end in $U$ and the other end in $W$ by $E(U, W)$; thus, $E(U, W) := \{uw \in E \ : \ u \in U, \ w \in W\}$.

We say that a link $uv$ *covers* a tree-edge $\hat{e}$ if $P_{u,v} \ni \hat{e}$.

For any tree-edge $\hat{e} \in \widehat{E}_T$, we use $\delta_E(\hat{e})$ to denote the set of links that cover $\hat{e}$, thus, $\delta_E(\hat{e}) = \{uv \in E : \ \hat{e} \in P_{u,v}\}$. For any node $w \in V$, we use $\delta_E(w)$ to denote the set of links incident to $w$.

**Shadows and the shadow-closed property.** For two links $u_1v_1$ and $u_2v_2$, if $P_{u_1,v_1} \subseteq P_{u_2,v_2}$, then we call $u_1v_1$ a *shadow* (or, *sublink*) of $u_2v_2$. In particular, if $P_{u_1,v_1} \subsetneq P_{u_2,v_2}$, then we call $u_1v_1$ a *proper shadow* (or, *proper sublink*) of $u_2v_2$.

For each link $uv \in E$, if all sublinks of $uv$ exist in $E$, then we call $E$ *shadow-closed*. Clearly, if $E$ is not shadow-closed, then we can make it shadow-closed by adding all sublinks of each of the original links. It can be seen that this preserves the optimal value of any instance of TAP.

Observe that a twin link cannot be a proper sublink of another link, because each end $u$ of a twin link has $\deg_T(u) = 1$. Thus, when we add sublinks to $E$ to make it shadow-closed, then none of the added sublinks can be a twin link. Hence, if we start with a stemless instance of TAP, then we do not introduce any stems when we make it shadow-closed.

Following Even, at al., see [9, Assumption 2.2], we make the next assumption.
**Assumption:** $E$ is shadow-closed.

**Contractions, compound nodes and original nodes.** By an *original node* we mean an element of $V(T)$, and by an *original link* we mean an element of $E$.

Recall that the algorithm (incrementally) constructs a set of links $F$ such that, at termination, $T + F = (V, \widehat{E}_T \cup F)$ is 2-edge-connected. Throughout, we use $F$ to denote the current solution of the algorithm; initially, $F := \emptyset$. We use the standard notion of *contracting* a link or a set of links, see [9], [4, Chapter 1]. Throughout, we use $T/F$ (or, $T' := T/F$) to denote the "current tree" obtained by contracting each of the 2-edge-connected components of $T + F = (V, \widehat{E}_T \cup F)$ to a single node. Each of the contracted nodes of $T/F$ is called a *compound node*, see [9, Section 3.2]; thus, each compound node corresponds to a set of two or more nodes of $V(T)$. Each of the other (non compound) nodes of $T'$ is called an *original node*. For this paragraph, given an original node $v$ of $T$, let us use $v'$ to denote the corresponding node of $T/F$, i.e., if $v$ is an original node of $T/F$, then we have $v' = v$, otherwise $v'$ denotes the compound node of $T/F$ that contains $v$. Similarly, let us use $u'w'$ to denote the "image" of an original link $uw \in E$ w.r.t. the current tree $T/F$. For a set of original links $J \subseteq E$, the image w.r.t. the current tree $T/F$ is $\{u'w' \,:\, uw \in J, u' \neq w'\}$. Note that original links that have both ends in the same compound node of $T/F$ are discarded since they are not relevant for the rest of the execution/analysis. When we discuss the algorithm and its analysis (in Sections 6, 7), we may abuse the notation by not distinguishing between an original link $uw \in E$ and its image $u'w'$ w.r.t. the current tree $T/F$; similarly, we may not distinguish between a set of original links and the set of images of those original links.

**Root, ancestor, descendant and rooted subtrees.** One of the nodes $r$ of $T$ is designated as the root; thus, we have a rooted tree $(T, r)$.

Let $v$ be a node of $T$. If a node $w$ belongs to the path $P_{v,r}$, then $w$ is called an *ancestor* of $v$, and $v$ is called a *descendant* of $w$. If a descendant $w$ of $v$ is adjacent to $v$ (thus, $w \neq v$), then $w$ is called a *child* of $v$, and $v$ is called a *parent* of $w$. Clearly, every node (except $r$) has a unique parent. If $v$ has no child, then we call $v$ a *leaf*; clearly, if $v$ has no child, then $\deg_T(v) = 1$. Note that $r$ is not a leaf, even if $\deg_T(r) = 1$. Throughout, we use $L$ to denote the set of original leaves; we use $\mathcal{R}$ to denote $V - L$, i.e., the set of original non-leaf nodes.

For any node $v$, we use $T_v$ to denote the rooted subtree of $(T, r)$ induced by $v$ and its descendants. (Throughout, we view $T$ as an "oriented tree" rooted at $r$, and we use the term subtree to refer to a rooted subtree.)

We say that a subtree $T_v$ is *covered* by a set of links $J \subseteq E$ if each tree-edge of $T_v$ is covered by some link of $J$.

**Property 2.1.** *Suppose $\bar{T}$ is a rooted tree. Let $\bar{T}_{v_1}$ and $\bar{T}_{v_2}$ be two (rooted) subtrees of $\bar{T}$. Then $\bar{T}_{v_1}$ and $\bar{T}_{v_2}$ either share no node or one is contained in the other.*

*Proof.* Suppose $\bar{T}_{v_1}$ and $\bar{T}_{v_2}$ share a node, say $w$. Then, both $v_1$ and $v_2$ are ancestors of $w$. This implies that one of $v_1$ or $v_2$ must be an ancestor of the other one. Hence, one of the two subtrees $\bar{T}_{v_1}$, $\bar{T}_{v_2}$ is contained in the other one (possibly, $v_1 = v_2$ and $\bar{T}_{v_1} = \bar{T}_{v_2}$). □

For any leaf $v$ of $T$, $up(v)$ denotes a node $q$ in $P_{v,r}$ that is nearest to the root and adjacent to $v$ via a link; clearly, $up(v)$ exists by the assumptions of feasibility and shadow-closure.

**Vectors and convex combinations.** For any vector $x \in \mathbb{R}^E$, let $ones(x)$ denote the set of links of $x$-value one, thus $ones(x) = \{uv \in E : x_{uv} = 1\}$.

For a vector $x \in \mathbb{R}^E$ and any subset $J$ of $E$, $x(J)$ denotes $\sum_{e \in J} x_e$, and $x|_J$ denotes the restriction of $x$ to $J$. Given several vectors $v^1, v^2, \ldots$, we write one of their convex combinations as $\sum_{i \in Z} \lambda_i v^i$; thus, $Z$ is a set of indices, and we have $\lambda_i \geq 0, \forall i \in Z$, and $\sum_{i \in Z} \lambda_i = 1$.

**Remark**. For expository reasons, Part I uses definitions of "stem" and "twin link" that are relaxations of the definitions in Part II. The two definitions in Part I are independent of the choice of the root $r$. In Part II, we call a node $s$ of the rooted tree $(T, r)$ a stem if $s \neq r$, $s$ has exactly two children, the subtree $T_s$ has exactly two leaves, and there exists a link between the two leaves; the leaf-to-leaf link is called a twin link. If $s$ is a stem according to Part II, then it must be a stem according to Part I, but not vice-versa; a similar statement holds for twin links. Clearly, if an instance of TAP is stemless according to Part I, then it is stemless according to Part II as well.

## 3. The initial LP

This section presents our LP ($LP_0$) for TAP.

Let $u_1v_1$ and $u_2v_2$ be a pair of links. We call it an *overlapping pair* of links if (i) $P_{u_1,v_1}, P_{u_2,v_2}$ have one or more tree-edges in common, and (ii) either an end of $u_1v_1$ is in $P_{u_2,v_2}$, or an end of $u_2v_2$ is in $P_{u_1,v_1}$. We call a set of links $J$ an *overlapping clique* if every pair of links in $J$ is an overlapping pair.

The following LP with constraints on overlapping pairs gives a formulation of shadow-closed instances of TAP.

$$(\boldsymbol{LP}_0) \quad \text{minimize}: \sum_{uv \in E} x_{uv}$$

$$\text{subject to}: \sum_{uv \in \delta_E(\hat{e})} x_{uv} \geq 1 \qquad \forall \hat{e} \in \widehat{E}_T$$

$$x_{u_1v_1} + x_{u_2v_2} \leq 1 \quad \forall \text{ overlapping pairs} \quad u_1v_1, u_2v_2 \in E$$

$$0 \leq x \leq 1$$

Note that it tightens the feasible region of the obvious *covering LP* for TAP,

$$\min\{\sum_{e \in E} x_e \ : \ x(\delta_E(\hat{e})) \geq 1, \forall \hat{e} \in \widehat{E}_T, \ 1 \geq x \geq 0\},$$

because ($LP_0$) has additional constraints for the overlapping pairs. The next result shows that ($LP_0$) has the same optimal value as the covering LP, for fractional solutions as well as for integral solutions, provided the instance is shadow-closed.

**Proposition 3.1.** *Consider a shadow-closed instance of TAP. The optimal values of* ($LP_0$) *and the covering LP are the same. Moreover, the best objective value of an integral solution of* ($LP_0$) *is the same as the best objective value of an integral solution of the covering LP.*

*Proof.* Clearly, any feasible solution of ($LP_0$) is also a feasible solution of the covering LP. We claim that there exists an optimal solution of the covering LP that is feasible for ($LP_0$). The first statement follows from this claim.

Let $x$ be an optimal solution for the covering LP that minimizes $\sum_{uv \in E} \mathit{length}(P_{u,v}) \cdot x_{uv}$, where $\mathit{length}(P_{u,v})$ denotes the number of tree-edges of $P_{u,v}$. We show that $x$ is feasible for $(LP_0)$. Otherwise, suppose that $x$ violates the constraint for an overlapping pair $u_1v_1, u_2v_2$. W.l.o.g., suppose that $u_1 \in V(P_{u_2,v_2})$ and some tree-edge is in both $P_{u_1,v_1}$ and $P_{u_2,v_2}$. Then, there is a maximal (nonempty) prefix of the edge sequence of $P_{u_1,v_1}$ that is contained in $P_{u_2,v_2}$; let us denote this prefix by $P_{u_1,u_*}$. (Note that the link $u_*v_1$ is present because the instance is shadow-closed.)

Let $\alpha$ denote the original value of $x_{u_1v_1}$. Then, we replace the value $x_{u_1v_1}$ by $1 - x_{u_2v_2}$, thereby enforcing the constraint for the overlapping pair $u_1v_1, u_2v_2$. Moreover, we add $\alpha + x_{u_2v_2} - 1$ to the value of $x_{u_*v_1}$, and if the new value of $x_{u_*v_1}$ exceeds 1, then we replace it by 1. It can be seen that this preserves the constraints of the covering LP. This procedure decreases the value $\sum_{uv \in E} \mathit{length}(P_{u,v}) \cdot x_{uv}$ but does not increase the objective value $\sum_{uv \in E} x_{uv}$. This contradicts the assumption that $x$ is an optimal solution for the covering LP that minimizes $\sum_{uv \in E} \mathit{length}(P_{u,v}) \cdot x_{uv}$.

The last part (on integral solutions) follows from similar arguments, because the above procedure maintains integrality. □

## 4. Lasserre tightening and its properties

We give a brief overview of the Lasserre system, [16]. In fact, our results can be stated and proved without going into the formalities of the Lasserre system. Essentially, we apply one well-known result about the Lasserre system, namely, the decomposition theorem of Karlin-Mathieu-Nguyen, see [12, 18, 19]. The comprehensive recent survey by Rothvoß [19] presents this result and much more.

Let $A$ be an $m \times n$ matrix, and let $(LP) \quad \min\{c^T x \ s.t. \ Ax \geq b\}$ be a linear programming relaxation of a binary integer programming problem; thus, each variable $x_1, \ldots, x_n$ is binary. We use the notation $A = (A_{\ell i})_{\ell \in [m],\, i \in [n]}$. (Below, we may use $(LP)$ to denote both the linear program above, and its feasible region; thus $LP = \{x \in \mathbb{R}^n \ : \ Ax \geq b\}$.)

An $n \times n$ matrix $M$ is called positive semidefinite (p.s.d.) if $x^T M x \geq 0, \ \forall x \in \mathbb{R}^n$; $M \succcurlyeq 0$ denotes that $M$ is p.s.d.

For a positive integer $t$ and the ground set $[n] = \{1, \ldots, n\}$, let $\mathcal{P}_t$ denote the family of subsets of $[n]$ of size at most $t$, i.e., $\mathcal{P}_t = \{I : I \subseteq [n], |I| \leq t\}$; thus, each element of $\mathcal{P}_t$ is an "index set" of size $\leq t$. In this subsection, we use $I$ and $J$ to denote elements of $\mathcal{P}_t$ (whereas, in the rest of the paper, we use $J$ to denote an arbitrary subset of $E(G)$).

The $t$-th level of the Lasserre system (or, hierarchy), denoted $\textsc{Las}_t(LP)$ consists of the vectors $y \in \mathbb{R}^{2^{[n]}}$ that satisfy

$$M_{t+1}(y) := (y_{I \cup J})_{I,J \in \mathcal{P}_{t+1}} \succcurlyeq 0; \quad \hat{M}_t^{\ell}(y) \succcurlyeq 0, \ \forall \ell \in [m]; \quad y_{\emptyset} = 1,$$

where $\hat{M}_t^{\ell}(y) := \left( \sum_{i=1}^{n} (A_{\ell i})(y_{I \cup J \cup \{i\}}) - (b_\ell)(y_{I \cup J}) \right)_{I,J \in \mathcal{P}_t}$.

In other words, the $(t+1)$-th level *moment matrix* of $y$, $M_{t+1}(y)$, is required to be p.s.d., and moreover, for each of the constraints of $(LP)$, namely, $\mathrm{row}_\ell(A)x - b_\ell \geq 0, \ \ell \in [m]$, the associated $t$-th level *moment matrix of slacks*, $\hat{M}_t^{\ell}(y)$, is required to be p.s.d.

The index $t$ of each relaxation in the sequence of tightened relaxations is known as the *level*; the level of the original relaxation is defined to be zero. For any $t = O(1)$, it is known that the relaxation at level $t$ can be solved to optimality (up to a "small enough" additive error term) in polynomial time, assuming that the original relaxation has polynomial size; additional mild conditions may be needed.

Moreover, the relaxation at level $n$ is exact.

Now, consider our LP ($LP_0$) for TAP, and let $\textsc{Las}_t(LP_0)$ denote the level $t$ tightening of ($LP_0$) by the Lasserre system. [‡] Rothvoß, see [18, Theorem 2], formulated the following decomposition theorem for the Lasserre system, based on an earlier decomposition theorem due to Karlin-Mathieu-Nguyen [12]. (We use this particular formulation and not the original statement of [12]; hence, we reference both [12] and [18].)

**Theorem 4.1.** *Let $J \subseteq E$. Let $k$ be a positive integer such that $|ones(x) \cap J| \leq k$ for every feasible solution $x$ of ($LP_0$). Then for every feasible solution $y \in \textsc{Las}_t(LP_0)$, where $t \geq k+1$, $y$ can be written as a convex combination: $y = \sum_{i \in Z} \lambda_i x^i$ such that $x^i$ is in $\textsc{Las}_{t-k}(LP_0)$ and $x^i|_J$ is integral (i.e., $x^i_{uv}$ is integral for each $uv \in J$), for all $i \in Z$.*

**Lemma 4.2.** *Let $J \subseteq E$ be an overlapping clique. For every feasible solution $x$ of ($LP_0$), we have $|ones(x) \cap J| \leq 1$. Furthermore, for every level $t \geq 2$ and every feasible solution $y$ of $\textsc{Las}_t(LP_0)$, we have $y(J) \leq 1$.*

*Proof.* Let $x$ be a feasible solution of ($LP_0$). Then the overlapping constraints in ($LP_0$) imply that $|ones(x) \cap J| < 2$. To see this, suppose that $|ones(x) \cap J| \geq 2$. Then there exists a pair of links $u_1v_1, u_2v_2 \in J$, with $x_{u_1v_1} = x_{u_2v_2} = 1$; thus, $u_1v_1, u_2v_2$ is an overlapping pair whose associated constraint in ($LP_0$) is violated.

By Theorem 4.1, $y$ can be written as a convex combination: $y = \sum_{i \in Z} \lambda_i x^i$ such that $x^i$ is in $\textsc{Las}_1(LP_0)$ and $x^i|_J$ is integral for each $i \in Z$. Hence, $x^i(J) \leq 1$ for each $i \in Z$, because $|ones(x^i) \cap J| \leq 1$ and $x^i|_J$ is integral. Consequently, the convex combination $y$ of $x^i, i \in Z$, satisfies $y(J) \leq 1$. □

**Lemma 4.3.** *Let $w$ be a leaf of $T$, and let $u$ be an ancestor of $w$ such that every internal node (if any) of $P_{w,u}$ has exactly one child. Let $\hat{e} = vu$ be the tree-edge of $P_{w,u}$ that is incident to $u$ (possibly, $v = w$). Then, $\delta_E(\hat{e})$ is an overlapping clique. In particular, $\delta_E(w)$ is an overlapping clique. Moreover, we have $y(\delta_E(w)) = 1$ for any feasible solution $y \in \textsc{Las}_t(LP_0)$ where $t \geq 2$.*

*Proof.* Consider any two links $f_1q_1, f_2q_2 \in \delta_E(\hat{e})$. Clearly, each of the links $f_1q_1, f_2q_2$ must have an end in $P_{w,v}$. Suppose that $q_1, q_2$ are the ends in $P_{w,v}$, and w.l.o.g., assume that $q_1$ is an ancestor of $q_2$. Then, observe that $f_1q_1, f_2q_2$ is an overlapping pair, because $q_1$ is in $P_{f_2,q_2}$ and the tree-edge $\hat{e}$ is in both $P_{f_1,q_1}$ and $P_{f_2,q_2}$ (see Figure 3). Hence, the set of links covering $\hat{e}$ is an overlapping clique.

If we take $\hat{e}$ to be the unique tree-edge incident to the leaf $w$, then it can be seen that $\delta_E(w) = \delta_E(\hat{e})$ is an overlapping clique. Moreover, by Lemma 4.2, $y(\delta_E(w)) \leq 1$, whereas the constraints of ($LP_0$) imply that $y(\delta_E(w)) = y(\delta_E(\hat{e})) \geq 1$. Therefore, $y(\delta_E(w)) = 1$. □

Recall that the matching polytope of the subgraph induced by the leaves, $G(L) = (L, E(L))$ is given by the following constraints:

$$x(\delta_{E(L)}(v)) \leq 1 \qquad \forall v \in L$$

$$x(E(W)) \leq \frac{|W|-1}{2} \qquad \forall W \subseteq L, |W| \text{ odd}$$

$$x \geq 0$$

The next result is essentially the result on the matching polytope from the survey of Rothvoß, see [19, Lemma 13, Sec 3.3], translated to our setting.

**Lemma 4.4.** *Let $\epsilon > 0$, and let $t \geq \frac{1}{2\epsilon} + 1$. Suppose that $y \in \textsc{Las}_t(LP_0)$ is a feasible solution to the $t$-th level of the Lasserre system. Then, $\frac{y|_{E(L)}}{1+\epsilon}$ is in the matching polytope of $G(L) = (L, E(L))$.*

[‡] Although we defined $\textsc{Las}_t(LP_0)$ to be a set of vectors in $\mathbb{R}^{2^{[|E|]}}$, in what follows, we abuse the notation and take $\textsc{Las}_t(LP_0)$ to be the projection on the subspace indexed by the singleton sets; thus, we take $\textsc{Las}_t(LP_0)$ to be a set of vectors in $\mathbb{R}^{|E|}$.

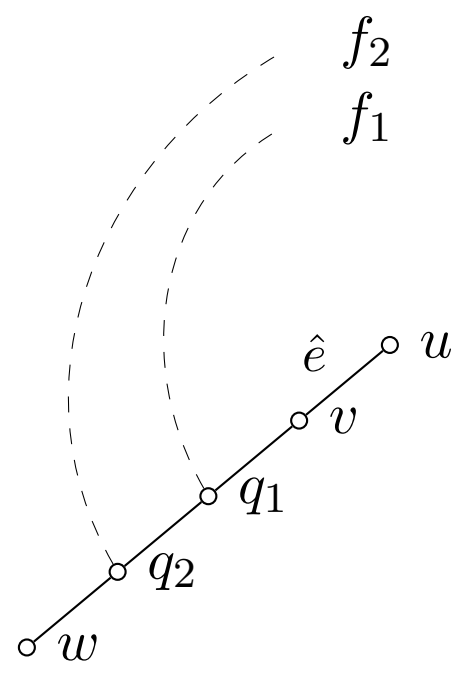

FIGURE 3. Illustration of the proof of Lemma 4.3. The solid lines are tree-edges and the dashed lines are links.

*Proof.* By Lemma 4.3, we have $y(\delta_{E(L)}(v)) \leq y(\delta_E(v)) \leq 1, \forall v \in L$. Hence, for any set $W \subseteq L$, we have $y(E(W)) \leq |W|/2$. For "large" odd sets $W \subseteq L$, we will show that $y(E(W)) \leq |W|/2$ implies that $\frac{y(E(W))}{(1+\epsilon)} \leq \frac{|W|-1}{2}$, whereas, for non-large odd sets $W$, we deduce $y(E(W)) \leq \frac{|W|-1}{2}$ by the decomposition theorem (Theorem 4.1) and local integrality on $E(W)$.

First, consider odd sets $W \subseteq L$ with $|W| > \frac{1}{\epsilon} + 1$. Clearly, we have $y(E(W)) \leq \frac{|W|}{2}$ since $y(\delta_{E(L)}(v)) \leq 1, \forall v \in L$. Also, observe that $\frac{|W|}{2} = \frac{|W|-1}{2}(1 + \frac{1}{|W|-1}) < \frac{|W|-1}{2}(1 + \frac{1}{\frac{1}{\epsilon}+1-1}) = \frac{|W|-1}{2}(1+\epsilon)$. Hence, $y(E(W)) \leq \frac{|W|-1}{2}(1+\epsilon)$.

Now, consider odd sets $W \subseteq L$ with $|W| \leq \frac{1}{\epsilon} + 1$. We apply the decomposition theorem, Theorem 4.1. Note that for any feasible solution $x$ of $(LP_0)$, by Lemmas 4.2, 4.3, we have $|ones(x) \cap E(W)| \leq \frac{|W|-1}{2} \leq \frac{1}{2\epsilon} \leq t-1$. Since $y$ is a feasible solution to the $t$-th level of the Lasserre system, $y$ can be written as a convex combination: $y = \sum_{i \in Z} \lambda_i x^i$ such that $x^i$ is in $\text{LAS}_1(LP_0)$ and $x^i|_{E(W)}$ is integral. Note that $\delta_E(w)$ is an overlapping clique by Lemma 4.3 for every $w \in W$. Hence, for each $i \in Z$, $x^i(E(W)) \leq \frac{|W|-1}{2}$. Consequently, $y(E(W)) \leq \frac{|W|-1}{2}$. This completes the proof. $\square$

## 5. POTENTIAL FUNCTION FOR STEMLESS TAP

This section presents the potential function that is used in our analysis.

Let $M$ denote a maximum matching of $(L, E(L))$; thus, $M$ is a maximum matching of the leaf-to-leaf links. By an $M$-link we mean a link that is in $M$. Let $U$ denote the set of $M$-exposed leaf nodes, that is, the set of leaves that are not covered by $M$.

We will often refer to $M$ and $U$ in the rest of the paper; these are key items for the algorithm (Section 6) and its analysis (Section 7).

Recall that $\mathcal{R}$ is the set of non-leaf nodes, i.e., $\mathcal{R} = V - L$.

**Lemma 5.1.** *Let $\epsilon > 0$ be a constant, and let $t \geq 1 + \frac{1}{2\epsilon}$. Let $y \in \mathbb{R}^E$ be any feasible solution of the $t$-th level of the Lasserre system, $\text{LAS}_t(LP_0)$. Then*

$$(\frac{3}{2} + \epsilon) y(E) \;\geq\; |U| + \frac{3}{2}|M| + \frac{1}{2}\sum_{v \in \mathcal{R}} y(\delta_E(v)).$$

*Proof.*

$$
\begin{aligned}
\frac{3}{2}y(E) &= \frac{3}{2}y(E(L)) + \frac{3}{2}y(E(L,\mathcal{R})) + \frac{3}{2}y(E(\mathcal{R})) \\
&\geq \Big(2y(E(L)) - \frac{1}{2}y(E(L))\Big) + \Big(y(E(L,\mathcal{R})) + \frac{1}{2}y(E(L,\mathcal{R}))\Big) + y(E(\mathcal{R})) \\
&= \Big(2y(E(L)) + y(E(L,\mathcal{R}))\Big) - \frac{1}{2}y(E(L)) + \Big(\frac{1}{2}y(E(L,\mathcal{R})) + y(E(\mathcal{R}))\Big) \\
&= \sum_{v\in L} y(\delta_E(v)) - \frac{1}{2}y(E(L)) + \frac{1}{2}\sum_{v\in\mathcal{R}} y(\delta_E(v)) \\
&\overset{(1)}{\geq} |L| - \frac{1}{2}(1+\epsilon)|M| + \frac{1}{2}\sum_{v\in\mathcal{R}} y(\delta_E(v)) \\
&\overset{(2)}{=} |U| + (\frac{3}{2} - \frac{\epsilon}{2})|M| + \frac{1}{2}\sum_{v\in\mathcal{R}} y(\delta_E(v)),
\end{aligned}
$$

where (1) follows from two facts that $y(\delta_E(v)) = 1$ for any $v \in L$ by Lemmas 4.3 and $\frac{y(E(L))}{1+\epsilon} \leq |M|$ by Lemma 4.4, and (2) follows from the observation that $|L| = |U| + 2|M|$. Note that $|M| \leq \frac{1}{2}|L| = \frac{1}{2}\sum_{v\in L} y(\delta_E(v)) \leq y(E)$, hence, $-\frac{\epsilon}{2}|M| \geq -\epsilon y(E)$. Thus, we have $(\frac{3}{2} + \epsilon)y(E) \ \geq\ |U| + \frac{3}{2}|M| + \frac{1}{2}\sum_{v\in\mathcal{R}} y(\delta_E(v))$. □

Lemma 5.1 states a key inequality for any feasible solution of the level-$t$ tightening of $(LP_0)$. But, the following example shows that this key inequality does not hold for all feasible solutions of $(LP_0)$ (without tightening by the Lasserre system). The example has a parameter $k$. We start with a path of length $k-1$, then attach a claw (a copy of $K_{1,3}$) at each node of the path. This gives the tree $T$. The link set $E$ consists of a link from one end of the path to the other end, and the three links connecting every pair of leaves in each copy of the claw. Note that all the sublinks of these links are also contained in $E$ (see Figure 4).

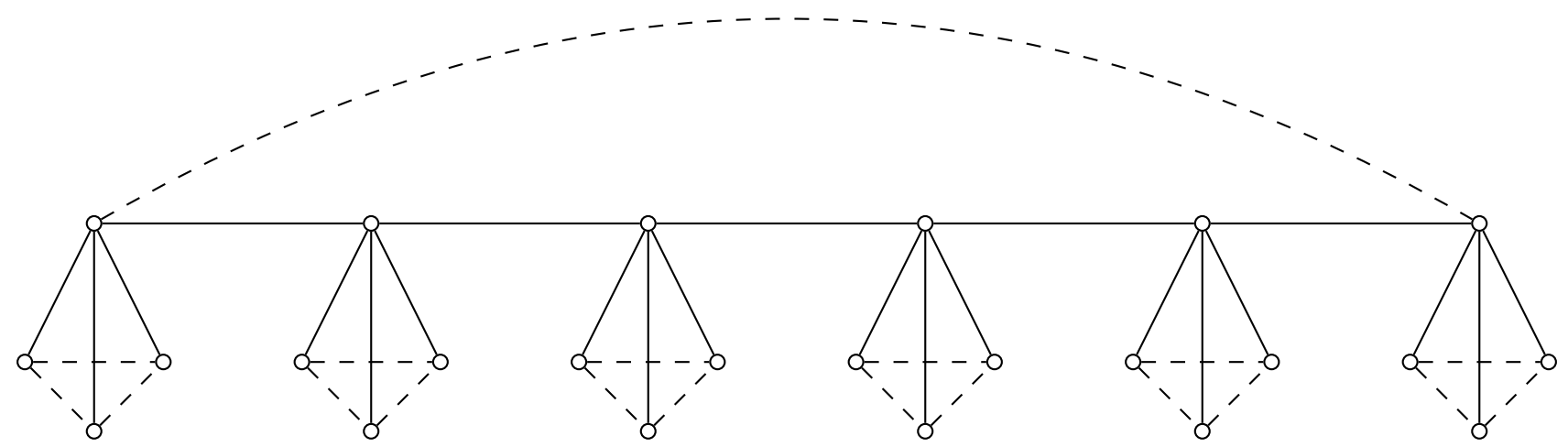

FIGURE 4. Instance for $k = 6$. The tree-edges of $T$ are indicated by solid lines, and the maximal links in $E$ are indicated by dashed lines.

Define a feasible solution $x$ of $(LP_0)$ as follows: the link from one end of the path to the other end gets value 1, every link connecting a pair of leaves in each claw gets value $\frac{1}{2}$, and all other links get value 0. It is not hard to see that $x$ is feasible for $(LP_0)$ and that $x(E) = \frac{3}{2}k + 1$. (Note that the optimal value of $(LP_0)$ may be $< x(E)$, but our arguments do not use the optimal value.)

If we pick any node on the path to be the root node, then any maximum matching $M \subseteq E(L)$ has size $k$ and there are $k$ $M$-exposed leaves. Thus, $\frac{3}{2}|M| + |U| = \frac{5}{2}k$, and this quantity is larger than $(\frac{3}{2}+\epsilon)x(E) = (\frac{9}{4}+\frac{3}{2}\epsilon)k+(\frac{3}{2}+\epsilon)$ for any $\epsilon < \frac{1}{6}$ and for sufficiently large $k$. Thus, the inequality stated in Lemma 5.1 does not hold for $x$.

## 6. Algorithm

This section presents the details of our algorithm and its analysis following the overview given in Section 1. The working of the algorithm is illustrated below, see Figure 5.

We first state our main result for (unweighted) stemless TAP:

**Theorem 6.1.** *Consider an instance of stemless TAP. Let $\epsilon > 0$ be a constant, and let $t \geq \max\{3,\ \frac{1}{2\epsilon}+1\}$. The integrality ratio of $\textsc{Las}_t(LP_0)$ is $\leq \frac{3}{2}+\epsilon$. Moreover, there is a polynomial-time algorithm for finding a feasible solution of TAP of size $\leq (\frac{3}{2}+\epsilon)y(E)$, where $y$ is an optimal solution of $\textsc{Las}_t(LP_0)$.*

Let $y \in \mathbb{R}^E$ be an optimal solution of the $t$-th level of the Lasserre system, $\textsc{Las}_t(LP_0)$, where $t \geq \max\{3,\ 1+\frac{1}{2\epsilon}\}$, where $\epsilon > 0$ is any (small) constant. We take the right-hand-side of the inequality in Lemma 5.1 to be our potential function. Thus, our potential function is $\leq (\frac{3}{2}+\epsilon)y(E)$. Our goal is to present an algorithm that finds a set of links $F$ that covers $T$ such that $|F|$ is $\leq$ our potential function; then, it will follow that $|F|$ is within a factor of $(\frac{3}{2}+\epsilon)$ of optimal.

The purpose of the potential function is to provide "credits" to the algorithm. In Section 6.2, we distribute the credit (i.e., the potential function) among the nodes and links.

Recall from Section 1 that the algorithm maintains a set of links $F$ and a current tree $T' := T/F$. Initially, we have $F := \emptyset$, and $T' := T$.

Consider $T' := T/F$ and the addition of a single link $\ell = vw$ to $F$. The new tree $T/(F \cup \{\ell\})$ can be obtained by contracting the unique path of $T'$ between $v$ and $w$, $P'_{v,w}$, to a single compound node. Besides adding a single link in a major step, the algorithm may add a set of links $\{\ell_1 = v_1w_1, \ell_2 = v_2w_2, \ldots, \ell_k = v_kw_k\}$ such that the union of the paths of $T'$ corresponding to $\ell_1, \ell_2, \ldots, \ell_k$ forms a single connected component, i.e., $\bigcup_{i=1}^{k} P'_{v_i,w_i}$ is a connected subgraph of $T'$. Again, the new tree can be obtained by contracting all of these paths into a single compound node.

The algorithm repeatedly finds a set of links $F^{iter} \subseteq E-F$ such that the contraction of $F^{iter}$ in the current tree results in a single new compound node, and moreover, the credit available from the contraction of $F^{iter}$ is $\geq |F^{iter}|+1$ (the details are discussed below). We add $F^{iter}$ to $F$, and obtain an updated current tree $T'$. The algorithm repeats this until $T'$ is a single node, that is, until $T+F$ is 2-edge-connected.

There are two types of link sets that get contracted by a major step. The first type is a singleton set, i.e., the major step adds one link to $F$. The second type is defined via the notion of a semiclosed tree. This is discussed in the next subsection.

### 6.1. Semiclosed trees.

We start by defining the key notion of a semiclosed tree w.r.t. an *arbitrary* matching of the leaf-to-leaf links. This notion is due to Even, et al., based on earlier work by Nagamochi [10]; also, see [9, Definition 2.3].

Let $T'_v$ be a rooted subtree of the current tree $T' = T/F$. Let $\bar{M}$ be an arbitrary matching of the leaf-to-leaf links. $T'_v$ is called *semiclosed* w.r.t. $\bar{M}$ if the following conditions hold:

- (i) Each link in $\bar{M}$ either has both ends in $T'_v$ or has no end in $T'_v$.
- (ii) Every link incident to an $\bar{M}$-exposed leaf of $T'_v$ has both ends in $T'_v$. (Thus, if $T'_v \neq T'$, then none of the links covering the tree-edge between $v$ and its parent is incident to an $\bar{M}$-exposed leaf of $T'_v$.)

Let $\bar{M}(T'_v)$ denote the set of links in $\bar{M}$ that have both ends in $T'_v$.

We define

$$\Gamma(\bar{M}, T'_v) \ :=\ \bar{M}(T'_v) \bigcup \{up(w)w\ :\ w \text{ is an } \bar{M}\text{-exposed leaf of } T'_v\};$$

thus, we associate a "basic link set" with the pair $\bar{M}, T'_v$. In general, the "basic link set" may not be a cover of $T'_v$.

By a *minimally semiclosed tree* $T'_v$ we mean that $T'_v$ is semiclosed but none of the proper rooted subtrees of $T'_v$ is semiclosed.

**Lemma 6.2** (Even, et al., [9])**.** *Let $T'_v$ be a minimally semiclosed tree w.r.t $\bar{M}$. Then $\Gamma(\bar{M}, T'_v)$ covers all the tree-edges of $T'_v$.*

*Proof.* (The proof is well known, but we include it for the sake of completeness.) Suppose that some tree-edge $\hat{e} = pq$ is not covered by $\Gamma(\bar{M}, T'_v)$, where $q$ is a child of $p$. Observe that $pq$ cannot be incident to a leaf of $T'_v$ since $\Gamma(\bar{M}, T'_v)$ has a link incident to each leaf of $T'_v$. Then, consider the rooted subtree $T'_q$. It can be seen that $T'_q$ is a semiclosed tree w.r.t. $\bar{M}$, otherwise, $\Gamma(\bar{M}, T'_v)$ would cover $\hat{e} = pq$. This contradicts the fact that $T'_v$ is minimally semiclosed. □

6.2. **Credit assignment.** Recall that $M$ is a maximum matching of $(L, E(L))$, $U$ is the set of $M$-exposed leaf nodes and $\mathcal{R}$ is the set of (original) non-leaf nodes in $T$.

We start with the credit given by the potential function of Lemma 5.1, and we maintain the following assignment of credits to the nodes of $T' := T/F$ and the links of $M$:

- every $M$-exposed original leaf has one credit,
- every compound node has one credit,
- every (original) node $v \in \mathcal{R}$ is assigned $\frac{1}{2}y(\delta_E(v))$ credit,
- every $M$-link has $\frac{3}{2}$ credit, and
- the (original) root $r$ has one credit.

It can be seen that the potential function of Lemma 5.1 suffices for assigning credits to the initial tree $T' := T$, except for the unit credit for the root $r$. When the algorithm terminates, the tree $T'$ becomes a single compound node with one credit. But, this credit will not be used any more, and thus we have a surplus of one credit. We assign this surplus credit to the root $r$ at initialization.

We mention that the nodes or links that get contracted into a compound node are no longer relevant for the algorithm or the analysis. In particular, the credit (if any) of such nodes or links may be used at the step when they get contracted into a compound node, but after that step, any remaining credit of such nodes or links is not used at all.

We take the credit of a link $\ell$ w.r.t. the current tree $T'$ to be the credit of the original link corresponding to $\ell$.

6.3. **Simple contractions and assertions on** $M$**.** Let $\ell = uw$ be a link, where $u, w$ are nodes of the current tree $T'$, and let $P'_{u,w}$ denote the path of $T'$ between $u$ and $w$. We call $\ell = uw$ a *good link* if the sum of the following items is $\geq 2$: (i) the credit of $\ell = uw$, (ii) the number of compound nodes in $P'_{u,w}$, (iii) the number of $M$-exposed original leaves in $P'_{u,w}$, and (iv) 1 if the root $r$ is an original node of $P'_{u,w}$. In other words, if we take the credits associated with $y$ to be "fractional credits," then $uw$ is a good link if the "non-fractional credits" associated with $uw$ and the nodes of $P'_{u,w}$ is $\geq 2$.

We define a *simple contraction* to be one of the following types of single-link contractions.

- For the current tree, consider a leaf-to-leaf link $uw$ such that each end owns one credit; thus each of $u, w$ is either a compound leaf node or an original leaf node that is $M$-exposed. Observe that $uw$ is a good link.
- For the current tree, consider an $M$-link $uw$ such that the path between $u$ and $w$ in the current tree contains at least one compound node. Again, note that $uw$ is a good link.

**Lemma 6.3** (Assertions on $M$)**.** *Suppose that no simple contractions are applicable. Then*

*(1) For every $M$-link $uw$, every node in the path between $u$ and $w$ in $T'$ is an original node. In particular, in $T'$, both ends of each $M$-link are original leaf nodes.*

*(2) There exist no links between $M$-exposed leaves.*

6.4. **Good semiclosed trees.** For the rest of the paper, unless mentioned otherwise, a semiclosed tree means a tree that is semiclosed w.r.t. the matching $M$.

Recall that a semiclosed tree is defined w.r.t. an arbitrary matching of the leaf-to-leaf links. We chose $M \subseteq E$ to be a maximum matching of the leaf-to-leaf (original) links of the (original) tree $T$. But, it is not obvious that the the "image" of $M$ in the current tree $T'$ is a matching of the leaf-to-leaf links of $T'$.

The image of $M$ w.r.t. $T'$ is $\{u'w' \ : \ uw \in M, u' \neq w'\}$. We abuse the notation and use $M$ to denote both $M$ and its image w.r.t. $T'$, and by an $M$-link of $T'$ we mean the image of an original $M$-link w.r.t. $T'$. Whenever we mention semiclosed trees w.r.t. $M$, we assume that no simple contractions (see Section 6.3) are applicable in the current tree $T'$. Then, Lemma 6.3(1) implies that $M$ is a set of leaf-to-leaf links w.r.t. the current tree $T'$. Hence, semiclosed trees w.r.t. $T'$ and $M$ are well defined.

Let $T'_v$ be a rooted subtree of $T'$. We use $M(T'_v)$ to denote the set of $M$-links of $T'$ that have both ends in $T'_v$. We use $U(T'_v)$ to denote the set of $M$-exposed leaves of $T'_v$, including both original leaf nodes and compound leaf nodes. Let $C(T'_v)$ denote the set of compound non-leaf nodes of $T'_v$. Moreover, for any vector $x \in \mathbb{R}^E$, we use $\Phi(x, T'_v)$ to denote $\frac{1}{2}\sum_{w \in V(T'_v)\cap\mathcal{R}} x(\delta_E(w))$. [§] We define the credit of $T'_v$ to be the sum of the credits of the nodes in $T'_v$ plus the sum of the credits of the links in $M(T'_v)$.

Observe that the credit of a semiclosed tree $T'_v$ is either $1+\frac{3}{2}|M(T'_v)|+|U(T'_v)|+|C(T'_v)|+\Phi(y, T'_v)$ (if $r \in V(T'_v)\cap\mathcal{R}$), or $\frac{3}{2}|M(T'_v)|+|U(T'_v)|+|C(T'_v)|+\Phi(y, T'_v)$ (if $r \notin V(T'_v)\cap\mathcal{R}$). We call a semiclosed tree $T'_v$ *good* if its credit is $\geq |\Gamma(M, T'_v)| + 1$.

**Lemma 6.4.** *Let $T'_v$ be a semiclosed tree. If at least one of the following conditions is satisfied, then $T'_v$ is good.*

- $T'_v = T'$
- $C(T'_v) \neq \emptyset$
- $|M(T'_v)| \geq 2$
- $\Phi(y, T'_v) \geq 1$
- $|M(T'_v)| = 1$ *and* $\Phi(y, T'_v) \geq \frac{1}{2}$.

*Proof.* First, suppose that the root $r$ is not in $V(T'_v) \cap \mathcal{R}$. This implies that if $T'_v = T'$, then $r$ must be contained in some compound non-leaf node, hence, $C(T'_v) \neq \emptyset$. Then, it can be seen that the difference between the credit of $T'_v$ and $|\Gamma(M, T'_v)| + 1$ is

$$\begin{aligned}
&= \frac{3}{2}|M(T'_v)| + |U(T'_v)| + |C(T'_v)| + \Phi(y, T'_v) - |\Gamma(M, T'_v)| - 1 \\
&= \frac{3}{2}|M(T'_v)| + |U(T'_v)| + |C(T'_v)| + \Phi(y, T'_v) - |M(T'_v)| - |U(T'_v)| - 1 \\
&= \frac{1}{2}|M(T'_v)| + |C(T'_v)| + \Phi(y, T'_v) - 1.
\end{aligned}$$

The above quantity is $\geq 0$ if any one of the conditions listed in the lemma is satisfied, hence, the result holds.

If the root $r$ is in $V(T'_v) \cap \mathcal{R}$, then $T'_v$ has one more unit of credit, and again it can be seen that the result holds. ☐

Consider a minimally semiclosed tree $T'_v$ and suppose that it is good. Then, by Lemma 6.2, $\Gamma(M, T'_v)$ is a cover of $T'_v$, and moreover, $T'_v$ has enough credit to pay for the contraction of $\Gamma(M, T'_v)$.

[§]Recall that $\mathcal{R}$ is the set of original non-leaf nodes of $T$; thus $V(T'_v) \cap \mathcal{R}$ denotes the set of nodes of $T'_v$ excluding all leaves and all compound nodes.

(Informally speaking, a naive algorithm could make progress whenever there exists a minimally semiclosed tree that is good.)

6.5. **Algorithm in summary.** We give a summary of the algorithm in pseudocode. The critical step of the algorithm is to find a good semiclosed tree and a cover of it of appropriate size in polynomial time. The details of this step are presented in Section 7. There, we show that if a semiclosed tree $T'_w$ is *not* good, then $T'_w$ and its incident links form a subgraph that we call a deficient 3-leaf tree (this is the key result of Part I). The algorithm finds all occurrences of deficient 3-leaf trees in polynomial time, and then computes an auxiliary matching of the leaf-to-leaf links of $T'$ that we denote by $M^{new}$. Then the algorithm finds a minimally semiclosed tree $T'_v$ w.r.t. $M^{new}$. We prove that $T'_v$ is good (it has enough credits to pay for the contraction of a cover of size $|\Gamma(M, T'_v)|$) and, moreover, $\Gamma(M^{new}, T'_v)$ is a cover of $T'_v$ of size $|\Gamma(M, T'_v)|$. (We mention that $M^{new}$ is an auxiliary matching that is used for finding a semiclosed tree $T'_v$ with desired properties, and other than that, the algorithm does not refer to $M^{new}$ at all; whereas, the matching $M$ and its image $M(T')$ are used throughout the algorithm and its analysis.)

The algorithm starts with $F := \emptyset$ ($F$ is the set of links picked by the algorithm) and $T' := T$ ($T'$ is the current tree $T/F$).

**while** $T'$ *is not a single node* **do**
    repeatedly apply simple contractions until no simple contractions are applicable;
    find a good semiclosed tree $T'_v$ with a cover $J$ of size $|\Gamma(M, T'_v)|$ (Algorithm 2 in Section 7 gives the details for finding such a good semiclosed tree) ;
    add $J$ to $F$, contract $T'_v$ to a new compound node, update $T'$;
**end**

**Algorithm 1:** Find an approximately optimal solution for TAP.

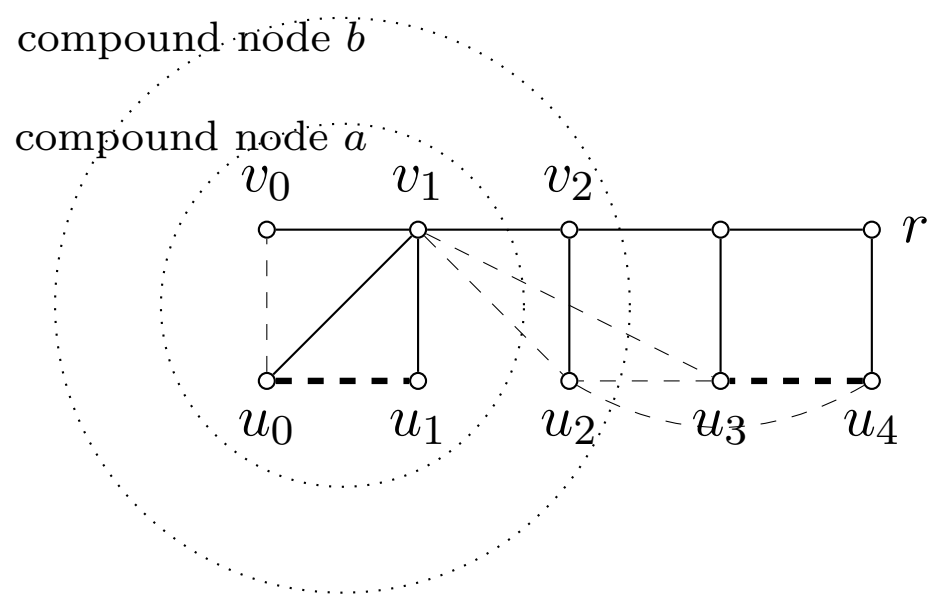

FIGURE 5. The edges of the tree $T$ rooted at $r$ are indicated by solid lines, and the maximal links in $E$ are indicated by dashed lines. The matching $M$ is indicated by the thick dashed lines.

6.6. **Worked example.** We give an informal presentation of the working of the algorithm, and illustrate it in Figure 5. (Section 8 discusses the working of the formal algorithm on an example.) Observe that $M$ consists of the two links $u_0u_1$, $u_3u_4$. Moreover, note that no simple contractions apply at the start, and the subtree rooted at $v_1$, $T'_{v_1} = T_{v_1}$, is a good semiclosed tree. In the first iteration (of the while loop), $T_{v_1}$ is contracted into the compound node $a$ by adding the $M$-link $u_0u_1$ and the link $v_0v_1$ (shadow of $v_0u_0$) to $F$; thus, $a$ corresponds to $T'_{v_1}$. Consider the credits for this iteration. The $M$-link $u_0u_1$ has $\frac{3}{2}$ credit and the $M$-exposed leaf $v_0$ has 1 credit. It can be seen that $\Phi(y, T'_{v_1}) \geq \frac{1}{2}y(\delta_E(v_1)) \geq \frac{1}{2}$, and this gives $\frac{1}{2}$ credit. (The formal algorithm does not refer to the "fractional credits" $\Phi(\cdot, \cdot)$, but our analysis relies on these; since we have not presented the formal algorithm in full, we refer to $\Phi(\cdot,\cdot)$ to justify the working of the algorithm.) Hence, we have $\geq 3$ credits, and this pays for contracting the two links and for assigning 1 credit to the compound node $a$. Next, the subtree rooted at $v_2$ (that has two leaves $a, u_2$) is contracted into the compound node $b$ via a simple contraction applied to the link $au_2$; thus, $b$ corresponds to the subtree rooted at $v_2$. Consider the credits for this simple contraction. Each of $a$ and $u_2$ has 1 credit, and these 2 credits pay for contracting one link and for assigning 1 credit to the compound node $b$. After this, the current tree $T'$ has three leaves, $b, u_3, u_4$. Note that $T'$ is a good semiclosed tree. In the final step, $T'$ is contracted into a single compound node by adding to $F$ the links $u_3u_4$ and $br$ (shadow of $bu_4$ corresponding to the original link $u_2u_4$). Consider the credits for the final step. Each of $b$ and $r$ has 1 credit, and the $M$-link $u_3u_4$ has $\frac{3}{2}$ credits. Hence, we have $3\frac{1}{2}$ credits, and this pays for contracting the two links and for assigning 1 credit to the resulting compound node. Thus, the algorithm computes the solution $F = \{v_0v_1, u_0u_1, au_2 = v_1u_2, br = u_2r, u_3u_4\}$ of size five; it can be seen that there exists an optimal solution of size four.

## 7. Analysis of the algorithm

This section has our main result. Informally speaking, it asserts the following: if a semiclosed tree $T'_v$ is not good, then $T'_v$ (and its incident links) form a deficient 3-leaf tree.

We assume that we are considering the moment after exhausting simple contractions in the main loop of the algorithm. Thus, Lemma 6.3 applies. The analysis mainly consists of two parts. In Section 7.1, using local integrality of feasible solutions to the Lasserre system, we show that all semiclosed trees are good, except one particular case that gives deficient 3-leaf trees. Section 7.2 shows how to handle deficient 3-leaf trees. This leads to a polynomial-time algorithm for finding a good semiclosed tree $T'_v$ and a cover of $T'_v$ of size $|\Gamma(M, T'_v)|$.

**Deficient 3-leaf trees.** Even, et al., introduced the notion of "deficient trees", see [9, Definition 4.7] and Figure 1 of [9]; each of these configurations consists of a rooted subtree with three leaves and some incident links.

Suppose that $T'_v$ is a semiclosed tree with exactly three leaves $a, b_1, b_2$. Clearly, among the nodes $w$ of $T'_v$ either there is exactly one node with $\deg_{T'}(w) = 4$ or there are two nodes with degree 3 in $T'$. In the latter case, we denote these two nodes by $u$ and $q$; moreover, we fix the notation such that $u$ is an ancestor of $q$, and the leaf $b_1$ is not a descendant of $q$; thus, $a, b_2$ (but not $b_1$) are descendants of $q$. In the former case, we denote by $u$ the unique node that is incident to four tree-edges. We call $T'_v$ a *deficient 3-leaf tree* if (i) the link $b_1b_2$ is present and it is in $M$, (ii) the link $ab_1$ is present, and (iii) there exists a link $b_2w$ such that $w \in V(T') - V(T'_v)$.

Moreover, in the first case (with a unique node $u$ in $T'_v$ with $\deg_{T'}(u) = 4$), if conditions (i)–(iii) hold with both labelings $(b_1, b_2)$ and $(b_2, b_1)$ of the $M$-link, then we fix the notation such that $up(b_2)$ is an ancestor of $up(b_1)$. We call $b_2$ the *ceiling leaf* of the deficient 3-leaf tree. Hence, for any deficient 3-leaf tree $T'_v$ with ceiling leaf $b$, it can be seen that $up(b)$ is a proper ancestor of $v$, i.e., $up(b)$ is not in $T'_v$.

We mention that the leaf $a$ may be an original node or a compound node; the properties of the algorithm (see Lemma 6.3) ensure that the leaves $b_1, b_2$ must be original nodes. Note that deficient 3-leaf trees are defined w.r.t. $M$.

### 7.1. Semiclosed trees are good except deficient 3-leaf trees.

Let $T'_v$ be a semiclosed tree. We construct an auxiliary graph (in fact, a multigraph, see below) in order to analyze the credits available in $T'_v$. We denote the auxiliary graph by $AG(T'_v)$. This is a bipartite (multi) graph, and the two sets in the node bipartition are denoted by $ML(T'_v)$ and $AU(T'_v)$. The first set consists of the $M$-covered leaves of $T'_v$. The second set contains an auxiliary node $\bar{v}$ (informally speaking, $\bar{v}$ represents the node set $V(T') - V(T'_v)$), as well as all the $M$-exposed leaves of $T'_v$, thus, $AU(T'_v) = \{\bar{v}\} \cup U(T'_v)$.

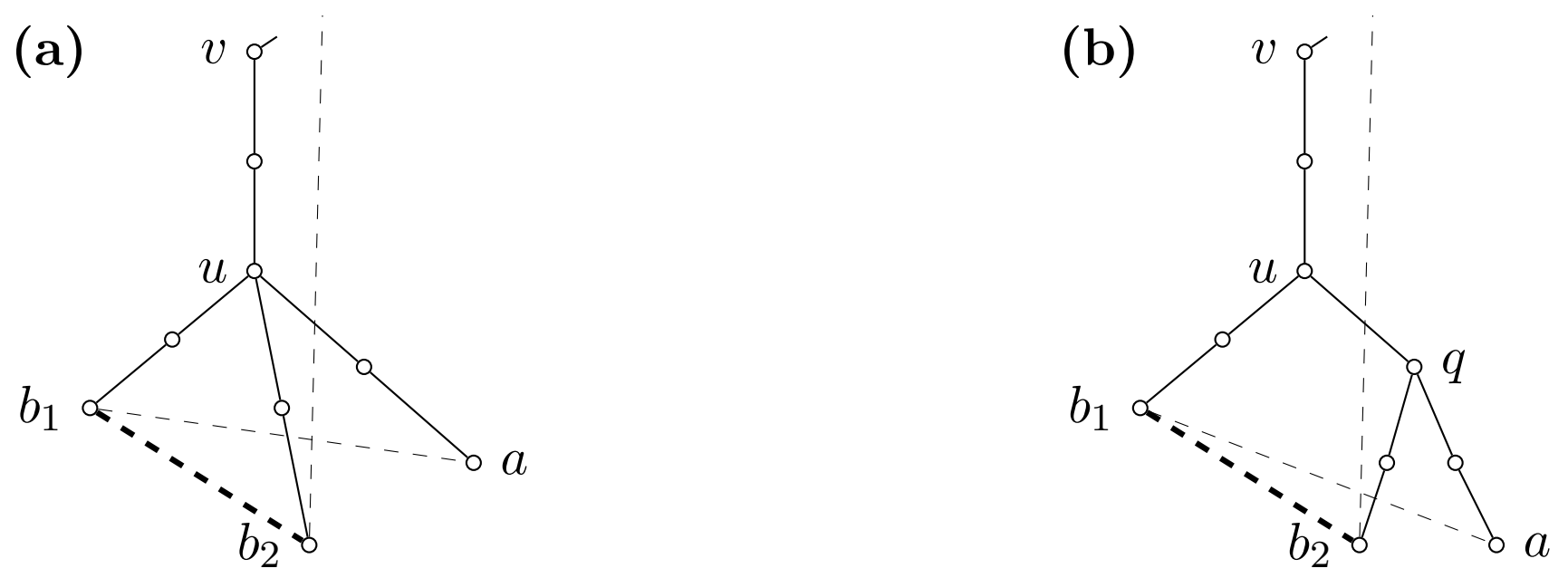

FIGURE 6. Illustration of deficient 3-leaf tree.

We define the edge set of $AG(T'_v)$ as follows: for every link $pq$ (w.r.t. $T'$) with $p \in ML(T'_v), q \in U(T'_v)$, the edge $pq$ is in $AG(T'_v)$, and for every link $pq$ (w.r.t. $T'$) with $p \in ML(T'_v), q \in V(T')-V(T'_v)$, the edge $p\bar{v}$ is in $AG(T'_v)$. Thus, $AG(T'_v)$ is a multigraph, and every edge in $AG(T'_v)$ corresponds to a link (w.r.t. $T'$). See Figure 7 for an example.

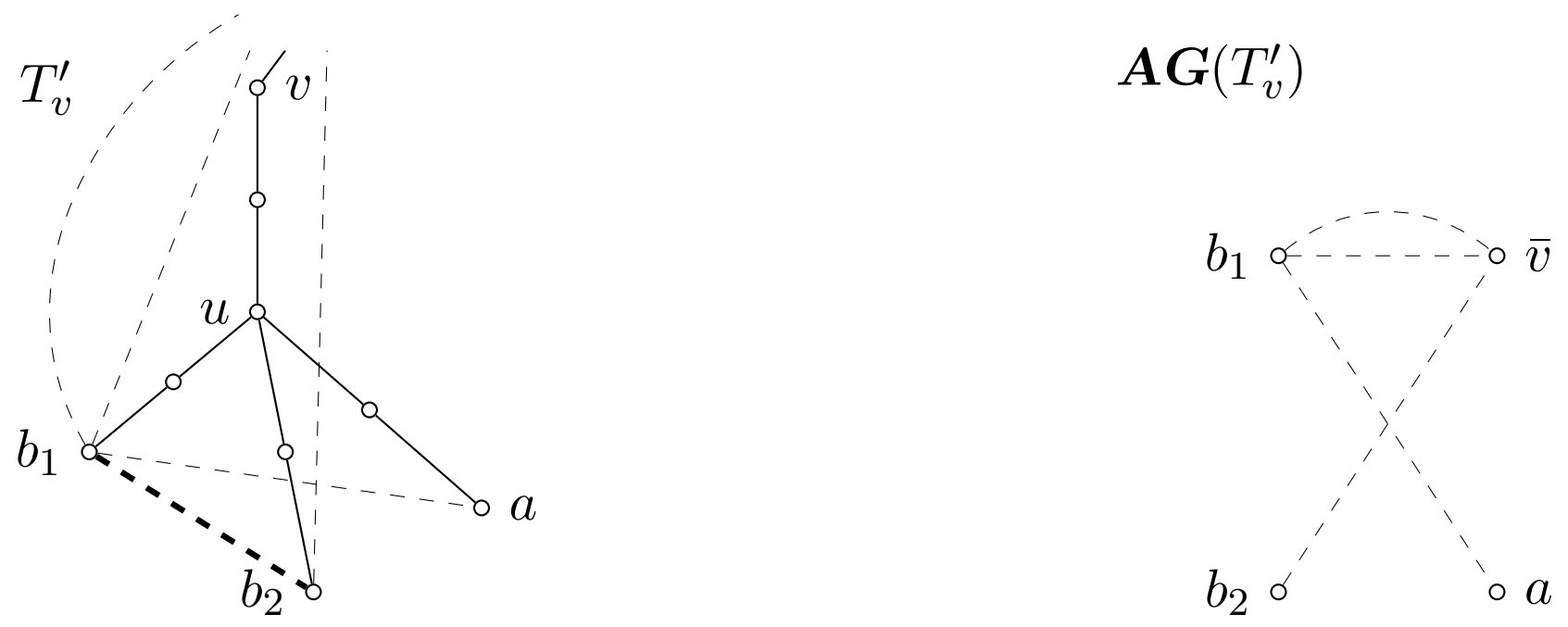

FIGURE 7. The left figure shows a semiclosed tree $T'_v$ where dashed lines indicate links incident with leaves and the thick dashed line indicates an $M$-link. The right figure shows the auxiliary graph $AG(T'_v)$ and its node bipartition $ML(T'_v) = \{b_1, b_2\}$, $AU(T'_v) = \{\bar{v}, a\}$.

**Lemma 7.1.** *Suppose that no simple contractions are applicable. Let $T'_v$ be a semiclosed tree such that $T'_v \neq T'$, $C(T'_v) = \emptyset$, and $|M(T'_v)| \leq 1$. Let $x$ be a feasible solution for ($LP_0$).*

*(1) If $M(T'_v) = \emptyset$, then $\Phi(x, T'_v) \geq 1$. Furthermore, $T'_v$ is good.*
*(2) Suppose that $|M(T'_v)| = 1$, and $|U(T'_v)| \geq 1$. Moreover, suppose that $x$ is integral on $\cup_{w \in ML(T'_v)} \delta_E(w)$ and that $\Phi(x, T'_v) < \frac{1}{2}$. Then, $|U(T'_v)| = 1$ and the auxiliary graph has a perfect matching such that the corresponding links cover (all the tree-edges in) $T'_v$, and moreover, $x(\ell) = 1$ for each of the links $\ell$ in the perfect matching.*

*Proof.* We start by stating and proving a key claim.

**Claim 7.2.** *Let $\bar{J}$ be a set of links that each have at least one end in $T'_v$ and no end at an $M$-covered leaf of $T'_v$; thus, each link in $\bar{J}$ has at least one end in $V(T'_v) - ML(T'_v)$. Then, we have $\Phi(x, T'_v) \geq \frac{1}{2} x(\bar{J})$.*

This claim follows from the fact that every link in $\bar{J}$ has an end in $V(T'_v) \cap \mathcal{R}$. To see this, first note that $C(T'_v) = \emptyset$, so every non-leaf node of $T'_v$ is in $V(T'_v) \cap \mathcal{R}$. Consider any link $\ell \in \bar{J}$; clearly, $\ell$ has no end at an $M$-covered leaf of $T'_v$. If $\ell$ has an end at a non-leaf node of $T'_v$, then we are done. Otherwise, $\ell$ has an end at an $M$-exposed leaf of $T'_v$. Since $T'_v$ is semiclosed, $\ell$ cannot have an end in $V(T')-V(T'_v)$. Moreover, by Lemma 6.3(2), no link has both ends at $M$-exposed leaves. It follows that $\ell$ has one end in $V(T'_v) \cap \mathcal{R}$. Thus, we proved Claim 7.2.

Let $\hat{e}_v$ denote the tree-edge between $v$ and its parent; $\hat{e}_v$ is well defined since $T'_v \neq T'$. Let $J = \delta_E(\hat{e}_v) \cup (\cup_{u \in U(T'_v)} \delta_E(u))$. Then, $x(J) = x(\delta_E(\hat{e}_v)) + \sum_{u \in U(T'_v)} x(\delta_E(u)) \geq 1 + |U(T'_v)|$; the equation holds because (i) $T'_v$ is semiclosed so none of the links in $\delta_E(\hat{e}_v)$ is incident to an $M$-exposed leaf of $T'_v$, and (ii) by Lemma 6.3(2), no link has both ends at $M$-exposed leaves; the inequality holds because $x(\delta_E(\hat{e})) \geq 1$ for every tree-edge $\hat{e}$.

Now, consider the first statement of the lemma. Observe that $M(T'_v) = \emptyset$; also, $U(T'_v) \neq \emptyset$ since $T'_v$ has one or more leaves. By Claim 7.2, we have $\Phi(x, T'_v) \geq \frac{1}{2}x(J) \geq \frac{1}{2}(1 + |U(T'_v)|) \geq 1$. Since this inequality holds for every feasible solution $x$ of $(LP_0)$, it also holds for the feasible solution $y$ of Theorem 6.1. Thus, we have $\Phi(y, T'_v) \geq 1$. Hence, by Lemma 6.4, $T'_v$ is good.

Finally, consider the second statement of the lemma. Observe that $|M(T'_v)| = 1$ and $|U(T'_v)| \geq 1$. Note that $x$ is integral on $\cup_{w \in ML(T'_v)} \delta_E(w)$, hence, every link in this set has $x$-value 0 or 1.

Moreover, by Lemma 6.3(1), every $M$-covered leaf is an original node. For every $w \in ML(T'_v)$, since $\delta_E(w)$ is an overlapping clique by Lemma 4.3 and $x$ is integral on $\delta_E(w)$, we have $x(\delta_E(w)) \leq 1$. Therefore, $\sum_{w \in ML(T'_v)} x(\delta_E(w)) \leq |ML(T'_v)| = 2$.

Let $\tilde{J} = J - \bigcup_{w \in ML(T'_v)} \delta_E(w)$. Then, Claim 7.2 holds for $\tilde{J}$, hence, $\Phi(x, T'_v) \geq \frac{1}{2}x(\tilde{J})$.

Note that $x(J) = x(\delta_E(\hat{e}_v)) + \sum_{u \in U(T'_v)} x(\delta_E(u)) \geq 1 + |U(T'_v)|$; this inequality is the same as the inequality used above. Moreover,

$$(1) \qquad x(\tilde{J}) \geq x(J) - \sum_{w \in ML(T'_v)} x(\delta_E(w) \cap J) \geq 1 + |U(T'_v)| - |ML(T'_v)| \geq |U(T'_v)| - 1.$$

Thus, we have $\Phi(x, T'_v) \geq \frac{1}{2}x(\tilde{J}) \geq \frac{1}{2}(|U(T'_v)| - 1)$.

Clearly, $|U(T'_v)| = 1$, otherwise, we would have $\Phi(x, T'_v) \geq \frac{1}{2}$, thus giving a contradiction. Hence, we have $|ML(T'_v)| = 2 = |AU(T'_v)|$.

Similarly, we claim that each $M$-covered leaf $w$ of $T'_v$ has a link $\ell_w$ in $\delta_E(w) \cap J$ of $x$-value one; otherwise, we would have $\sum_{w \in ML(T'_v)} x(\delta_E(w) \cap J) \leq |ML(T'_v)| - 1$ since every link in $\delta_E(w) \cap J$ takes $x$-value 0 or 1. This would give the same contradiction (see (1) above).

We claim that the set of links $\{\ell_w \ : \ w \in ML(T'_v)\}$ maps to the desired perfect matching $AM$ of $AG(T'_v)$. Otherwise, one of the nodes of $AU(T'_v)$ would be incident to two links from $\{\ell_w : w \in ML(T'_v)\}$, and we would have $x(J) = x(\delta_E(\hat{e}_v)) + \sum_{u \in U(T'_v)} x(\delta_E(u)) \geq 2 + |U(T'_v)|$, and this would give the same contradiction (see (1) above).

Finally, we claim that $T'_v$ is covered by the set of links $\{\ell_w \ : \ w \in ML(T'_v)\}$ that maps to $AM$. Otherwise, there exists a tree-edge $\hat{e}$ of $T'_v$ that is not covered by this set of links. Let $\delta^+(\hat{e})$ denote the set of links with positive $x$-value that cover $\hat{e}$. Clearly, each link in $\delta^+(\hat{e})$ has an end in $T'_v$, and moreover, has no end in $ML(T'_v)$; to see the latter assertion, note that each node $w \in ML(T'_v)$ has $x(\delta_E(w)) \leq 1$ and $x(\ell_w) = 1$, i.e., the nodes in $ML(T'_v)$ are already "saturated" by the set of links that maps to $AM$. Thus, Claim 7.2 applies to $\delta^+(\hat{e})$ and we have $\Phi(x, T'_v) \geq \frac{1}{2}x(\delta^+(\hat{e})) \geq \frac{1}{2}$, giving the same contradiction. □

**Theorem 7.3.** *Suppose that no simple contractions are applicable. Let $T'_v$ be a semiclosed tree that is not good. Then $T'_v$ is a deficient 3-leaf tree.*

*Proof.* Since $T'_v$ is not good, Lemma 6.4 implies that $C(T'_v) = \emptyset, |M(T'_v)| \leq 1$, and $T'_v \neq T'$. Then by Lemma 7.1(1), we further have $|M(T'_v)| = 1$. Hence, by Lemma 6.4 again, $\Phi(y, T'_v) < \frac{1}{2}$.

Let $J = \cup_{w \in ML(T'_v)} \delta_E(w)$. Note that any node $w$ in $ML(T'_v)$ is original, by Lemma 6.3(1). For any feasible solution $x \in \mathbb{R}^E$ of $(LP_0)$, we have $|ones(x) \cap J| \leq 2$ since $|ML(T'_v)| = 2$. Hence, by Theorem 4.1, and the fact that $t \geq 3$, $y$ can be written as a convex combination $\sum_{i \in Z} \lambda_i x^i$ such that $x^i \in \textsc{Las}_{t-2}(LP_0)$ and $x^i$ is integral on $J$ for each $i \in Z$. If each $x^i$, $i \in Z$, has $\Phi(x^i, T'_v) \geq \frac{1}{2}$, then, since $y$ is a convex combination of the $x^i$, we have $\Phi(y, T'_v) \geq \frac{1}{2}$. This gives a contradiction. Hence, there exists an $i_0 \in Z$ such that $\Phi(x^{i_0}, T'_v) < \frac{1}{2}$.

**Case (1):** $|M(T'_v)| = 1, U(T'_v) = \emptyset$. Let $uw \in M(T'_v)$, i.e., $uw$ is an $M$-link with both ends in $T'_v$. There exists no compound node in the path between $u$ and $w$ in $T'$, by Lemma 6.3. Then, it can be seen that $T'_v$ contains a stem with $uw$ as a twin link. This is a contradiction, because we have an instance of stemless TAP.

**Case (2):** $|M(T'_v)| = 1$, $|U(T'_v)| \geq 1$. By Lemma 7.1(2), we know $|U(T'_v)| = 1$. (For convenience, we will label the nodes of $T'_v$ using the same labels as in Figure 6 but our arguments do not rely on these particular labels.) We denote the $M$-exposed leaf by $a$, and the two $M$-covered leaves by $b_1, b_2$, i.e., $ML(T'_v) = \{b_1, b_2\}$. Now, our goal is to show that $T'_v$ satisfies all the conditions of a deficient 3-leaf tree.

Since $T'_v \neq T'$, let $\hat{e}$ denote the tree-edge between $v$ and its parent. By Lemma 7.1(2), there exist two links $\ell_v \in \delta_E(\hat{e})$ and $\ell_a \in \delta_E(a)$ such that $x^{i_0}(\ell_v) = x^{i_0}(\ell_a) = 1$, these two links cover $T'_v$, and moreover, each of $b_1, b_2$ is incident to exactly one of these two links (since the auxiliary graph has a perfect matching formed by these two links).

If there is only one non-leaf node $u$ in $T'_v$ with $\deg_{T'}(u) \neq 2$ (see Figure 6(a)), then we are done. Otherwise, we have exactly two non-leaf nodes $u, q$ in $T'_v$ with $\deg_{T'}(u) \neq 2$ and $\deg_{T'}(q) \neq 2$. In fact, we must have $\deg_{T'}(u) = 3 = \deg_{T'}(q)$ since $T'_v$ has exactly 3 leaves. W.l.o.g., we assume that $u$ is an ancestor of $q$. Then, $T'_q$ has only two leaves. By the argument in Case (1), the $M$-link in $T'_v$ cannot have its two ends at the two leaves of $T'_q$. This implies that one leaf of $T'_q$ is $M$-exposed; thus $a$ is a leaf of $T'_q$. W.l.o.g., take the other leaf of $T'_q$ to be $b_2$. Thus, the third leaf $b_1$ is not in $T'_q$.

Suppose that $\ell_v$ is incident to $b_1$ and $\ell_a$ is incident to $b_2$. Then, the tree-edge between $q$ and its parent is not covered by these two links (see Figure 8(a)). This is a contradiction. Hence, $\ell_v$ is incident to $b_2$ and $\ell_a$ is incident to $b_1$ (see Figure 8(b)). Therefore, $T'_v$ satisfies all the conditions of a deficient 3-leaf tree.

□

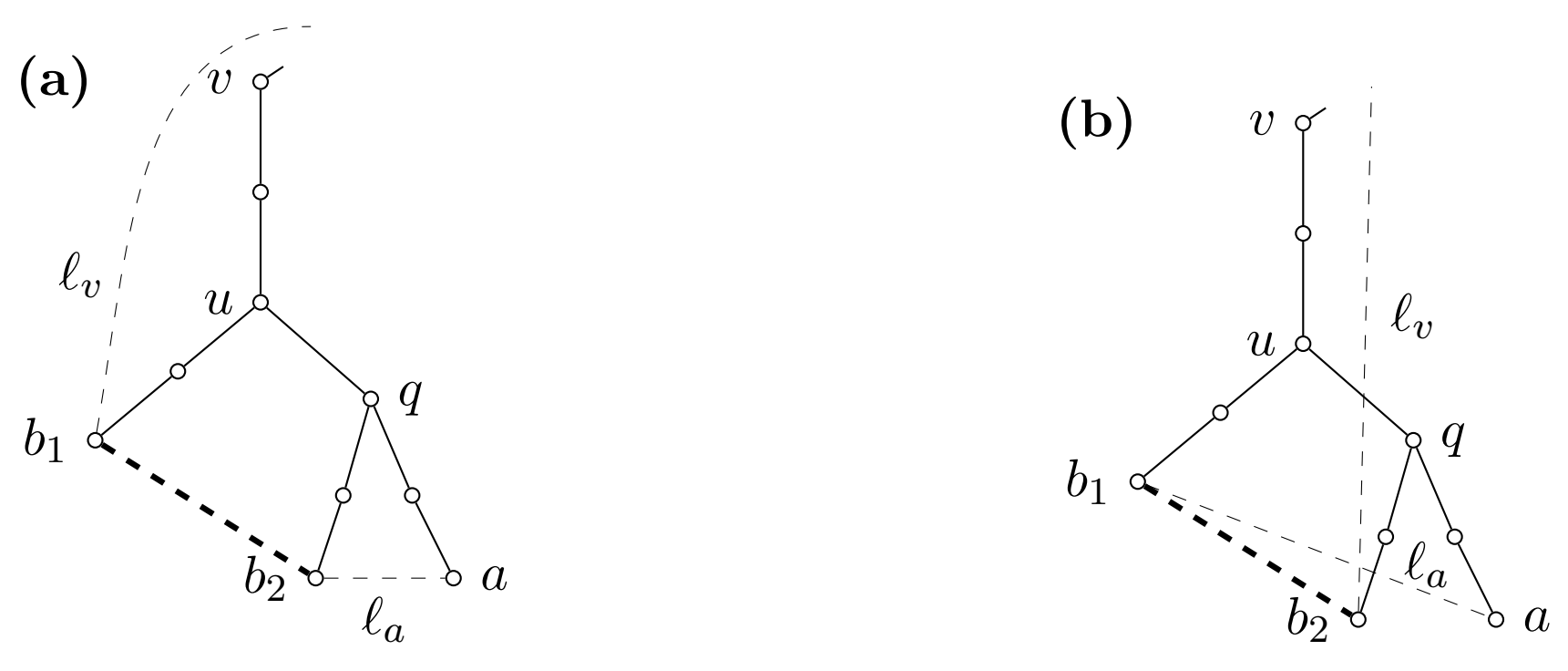

FIGURE 8. The links $\ell_v$ and $\ell_a$ in the proof of Theorem 7.3.

7.2. **Addressing deficient 3-leaf trees.** Even, et al., see [9, Section 4.3], presented an elegant method for addressing deficient 3-leaf trees. We use essentially the same method in this section. The key point is to compute an auxiliary matching of the leaf-to-leaf links of $T'$ denoted by $M^{new}$, and then to find a minimally semiclosed tree $T'_v$ w.r.t. $M^{new}$. (As mentioned in Section 6.5, $M^{new}$ and $M$ have different purposes, and $M$ stays the same throughout the execution.)

For a deficient 3-leaf tree $T'_w$, if $T'_w$ is not a proper subtree of another deficient 3-leaf tree, then we call $T'_w$ a *maximal deficient 3-leaf tree.* By Property 2.1, any two different maximal deficient 3-leaf trees are disjoint. To construct $M^{new}$, we start with $M^{new} := M$, then we examine each maximal deficient 3-leaf tree $T'_w$ and we replace the unique link of $M(T'_w)$ in $M^{new}$ by another leaf-to-leaf link. In more detail, consider any maximal deficient 3-leaf tree $T'_w$, and let the three leaves be $a, b, d$, where $a$ is $M$-exposed, $b$ is the ceiling leaf, and $bd$ is the unique link in $M(T'_w)$; we keep the link $ad$ in $M^{new}$ instead of the $M$-link $bd$ (see Figure 9). Since any two different maximal deficient 3-leaf trees are disjoint, this replacement takes place independently for each maximal deficient 3-leaf tree.

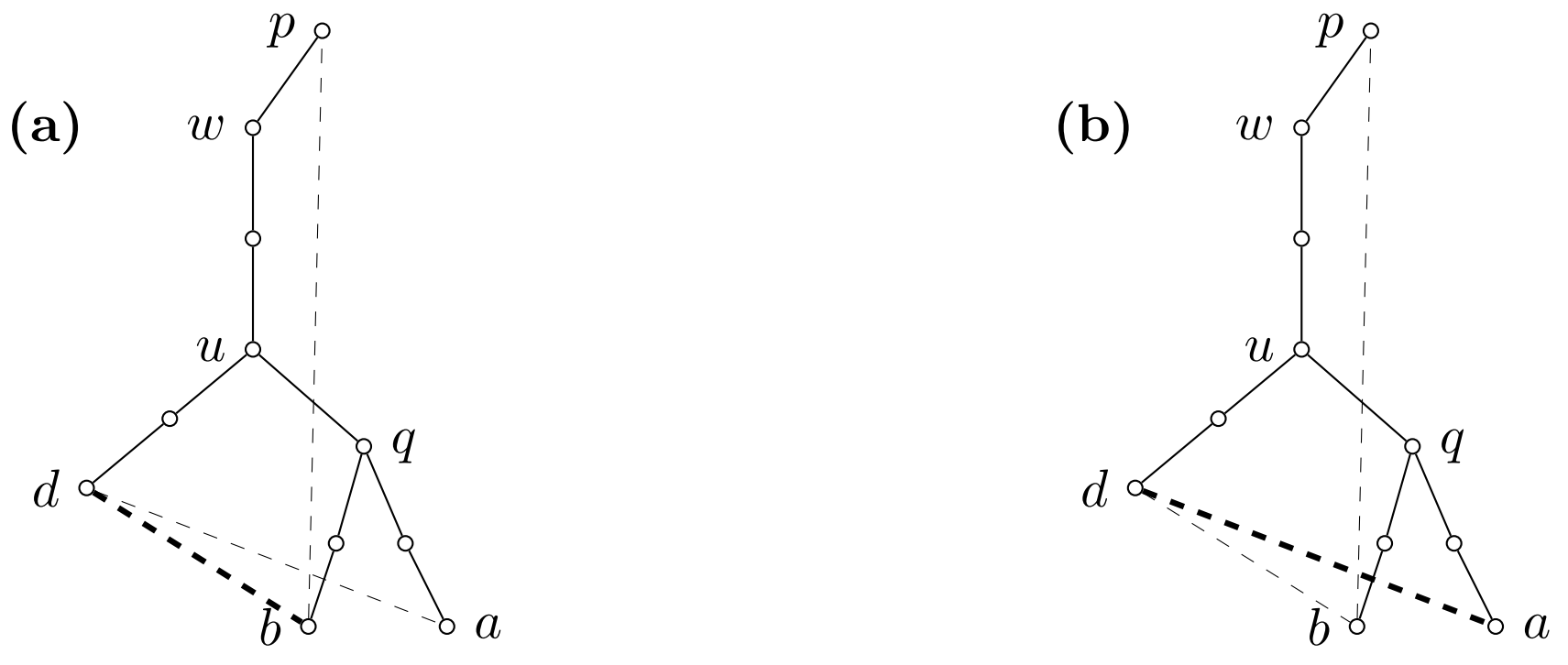

FIGURE 9. Addressing deficient 3-leaf trees by replacing $M$ by $M^{new}$. Figure (*a*) shows a maximal deficient 3-leaf tree $T'_w$ with ceiling leaf $b$ and $M$-exposed leaf $a$. We obtain the matching $M^{new}$ from $M$ by replacing the link $db$ by the link $da$; see Figure (*b*). $T'_w$ is not a semiclosed tree w.r.t. $M^{new}$ since $b$ is $M^{new}$-exposed. Instead, $T'_p$ is a minimally semiclosed tree w.r.t $M^{new}$. Note that $T'_p$ is not a deficient 3-leaf tree.

**Theorem 7.4.** *Suppose that no simple contractions are applicable. Let $T'_v$ be a minimally semiclosed tree w.r.t. $M^{new}$. Then $T'_v$ is a good semiclosed tree w.r.t. $M$ and $T'_v$ has a cover $\Gamma(M^{new}, T'_v)$ of size $|\Gamma(M, T'_v)|$.*

*Proof.* We start by stating and proving a key claim.

**Claim**. If $T'_v$ has a node in common with a maximal deficient 3-leaf tree, then $T'_v$ properly contains this maximal deficient 3-leaf tree.

Let $T'_v$ share a node with a maximal deficient 3-leaf tree $T'_w$; let $b$ denote the ceiling leaf of $T'_w$. Note that $up(b)$ is not in $T'_w$. Observe that $b$ is an $M^{new}$-exposed node. By the definition of deficient 3-leaf tree, no subtree of $T'_w$ is semiclosed w.r.t. $M^{new}$. This implies that $T'_v$ cannot be a subtree of $T'_w$. Then, by Property 2.1, $T'_v$ properly contains $T'_w$. This proves the claim.

The claim shows that either $T'_v$ is disjoint from every maximal deficient 3-leaf tree, or $T'_v$ properly contains some maximal deficient 3-leaf trees. Thus, $T'_v$ cannot be a deficient 3-leaf tree.

By the construction of $M^{new}$ and the fact that $T'_v$ is semiclosed w.r.t. $M^{new}$, it can be seen that $T'_v$ is semiclosed w.r.t. $M$. Moreover, it can be seen that $|M(T'_v)| = |M^{new}(T'_v)|$, and hence, we have $|\Gamma(M, T'_v)| = |\Gamma(M^{new}, T'_v)|$.

Since $T'_v$ is a minimally semiclosed tree w.r.t. $M^{new}$, Lemma 6.2 implies that $\Gamma(M^{new}, T'_v)$ is a cover of $T'_v$. We already showed that the size of $\Gamma(M^{new}, T'_v)$ is equal to $|\Gamma(M, T'_v)|$.

Finally, observe that $T'_v$ is semiclosed w.r.t. $M$ and $T'_v$ is not a deficient 3-leaf tree, hence, Theorem 7.3 implies that $T'_v$ is good. This completes the proof. $\square$

The procedure for finding a good semiclosed tree is summarized in the following pseudocode.

```
start with M^new := M;
for each maximal deficient 3-leaf tree T'_w do
    let b be the ceiling leaf, a be the M-exposed leaf, and db be the M-link in T'_w;
    update M^new by replacing db by da (M^new := M^new − {db} ∪ {da});
end
(note that M^new is a matching of the leaf-to-leaf links of T');
find a minimally semiclosed tree T'_v w.r.t. M^new ;
T'_v is a good semiclosed tree w.r.t. M with a cover Γ(M^new, T'_v) of size |Γ(M, T'_v)| by Theorem
7.4;
```

**Algorithm 2:** Find a good semiclosed tree by addressing all deficient 3-leaf trees.

The discussion above shows how to find a good semiclosed tree $T'_v$ with a cover of size $|\Gamma(M, T'_v)|$ in polynomial time, for the main loop in our algorithm in Section 6. Therefore, Algorithm 1 (the overall algorithm) runs in polynomial time and returns a solution for TAP with size $\leq (\frac{3}{2} + \epsilon)y(E)$. This proves Theorem 6.1.

It can be seen that Algorithm 1 (the overall algorithm) can be implemented to run in time $O(|V|^3)$, assuming that the number of (multi) links in the input is $O(|V|^2)$. This can be achieved using simple data structures such as adjacency matrices and adjacency lists; the details are not straightforward, but we skip them.

**Theorem 7.5.** *Algorithm 1 can be implemented to run in time $O(|V|^3)$.*

## 8. Tight example for the analysis

In this section, we present a tight example to show that the approximation guarantee of our algorithm cannot be improved beyond $\frac{3}{2}$. The example has a parameter $k$. It consists of one initial block and $k$ copies of a repeated block. See Figure 10.

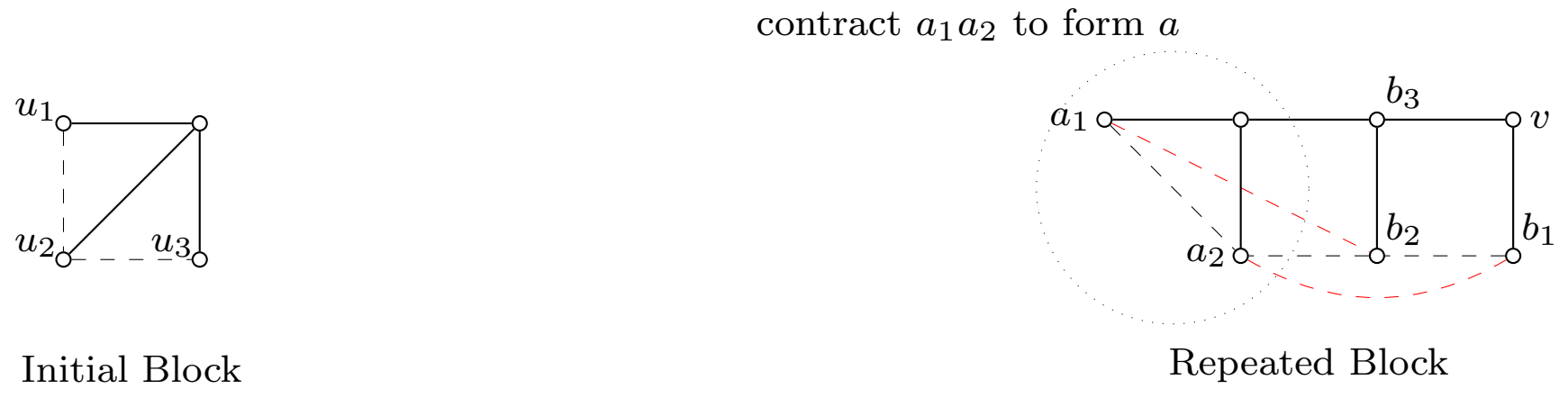

Figure 10. Two building blocks of our example. The maximal links are indicated by dashed lines.

Figure 11 shows an instance with $k = 3$ copies of the repeated block and all maximal links in $E$ are shown in the figure. Clearly, our instance has no stems. An instance for large $k$ can be constructed by adding more copies of the repeated block. The root node $r$ is always in the rightmost copy of the repeated block.

In our example, the blocks are disjoint in terms of both tree-edges and links. This implies that any feasible solution of TAP must cover each block individually. To find an optimal solution, we only need to consider each block individually. For the initial block, we need two links to cover the three tree-edges. For the repeated block, we again need two links to cover the six tree-edges (see the links $a_1b_2$, $a_2b_1$ in Figure 10). Hence, an optimal solution of our instance has size $2k + 2$.

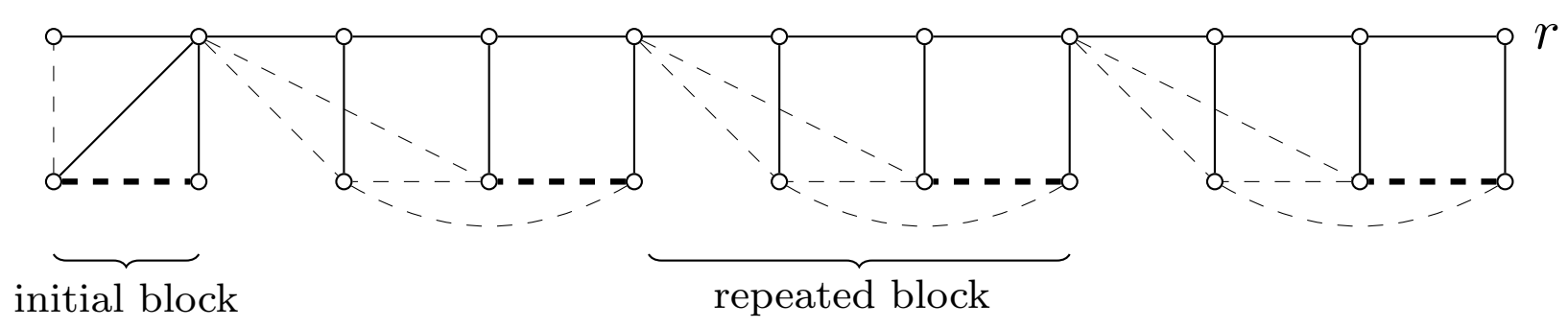

Figure 11. Instance for $k = 3$. The tree-edges of $T$ are indicated by solid lines, and the maximal links in $E$ are indicated by dashed lines. The maximum matching $M$ is indicated by thick dashed lines.

Now consider the execution of our algorithm on this instance. The maximum matching $M$ is shown in Figure 11. At the start, there is no good link available for simple contractions. The initial block is a minimally semiclosed tree (but not a deficient 3-leaf tree). So this block will be contracted via the M-link in it and the link from the $M$-exposed leaf to the root of the block. After that, we enter into the repeated block (see Figure 10). At this moment, the link $a_1a_2$ connects two $M$-exposed leaves, which implies it is a good link; note that there are no other good links. Hence, we apply a simple contraction on the link $a_1a_2$ to form a compound node $a$ (see Figure 10). Then, the repeated block forms a minimally semiclosed tree $T'_v$ with 3 leaves $a, b_1, b_2$ and one $M$-link $b_2b_1$ (see Figure 10). Note that $T'_v$ is not a deficient 3-leaf tree since there exists no link between an $M$-covered leaf of $T'_v$ and a node not in $T'_v$. Thus $T'_v$ is good. In the next step, $T'_v$ will be contracted via the two links $av$ (shadow of $ab_1$) and $b_1b_2$. After that, we enter into another repeated block, and we apply the same steps as for the previous repeated block. The algorithm applies these iterations till it terminates.

During the running of the algorithm, we use 2 links for the initial block and 3 links for each repeated block (one link for simple contraction and two links for contracting a good semiclosed tree with 3 leaves). Hence, the algorithm returns a solution of size $2+3k$. Therefore, the approximation guarantee of the algorithm is $\frac{2+3k}{2+2k}$. When $k$ is sufficiently large, the approximation guarantee approaches $\frac{3}{2}$. This shows that the approximation guarantee of our algorithm cannot be improved beyond $\frac{3}{2}$.

Note that the above instance of TAP has some cut nodes. But, we can modify the construction to get a 2-node-connected instance by adding some links to the above instance. We add a link from the leaf incident with both maximal links in the initial block (see node $u_2$ in Figure 10) to the non-leaf child of the root of the first repeated block (see node $b_3$ in Figure 10). Moreover, for each pair of consecutive repeated blocks, we add a link between the non-leaf children of their roots. It can be seen that the addition of these links does not change the working of the algorithm. Clearly, the addition of these links cannot increase the size of an optimal solution. It follows that the approximation guarantee of the algorithm is at least $\frac{2+3k}{2+2k}$, even for 2-node-connected instances.

**Proposition 8.1.** *The instance presented above shows that the algorithm in Section 6 cannot provide an approximation guarantee better than* $\frac{3}{2}$.

## 9. A preview of Part II

Our algorithm and analysis for TAP in Part II follow the same global scheme as in Part I, that is, we start by deriving a potential function, then we design an algorithm based on the same two types of contractions (greedy contractions and the contraction of good semiclosed trees), and we prove a similar key assertion (if a semiclosed tree is not good, then it has a particular configuration, namely, a deficient tree); the key assertion gives us the algorithmic tools for finding a good semiclosed tree.

Unfortunately, there are several "small configurations" that cannot be handled directly by this scheme, hence, the scheme has to be modified; both the potential function and the algorithm are modified, hence, our analysis becomes more complicated.

Let us illustrate some aspects of the cascade of difficulties by discussing the most obvious difficulty, namely, stems and twin links; let us apply the potential function (and credits) of Part I, that is, the right-hand side of the inequality in Lemma 5.1. Consider a stem $s$ and the subtree $T_s$; recall that $T_s$ has exactly two leaves and there is a so-called twin link $\ell_s$ between these two leaves. Suppose that the matching $M$ contains $\ell_s$; thus, $\ell_s$ is an $M$-link and it has $\frac{3}{2}$ credits. Moreover, suppose that no other credits are available in $T_s$ (this is possible, because all the links $\ell$ with $y(\ell) > 0$ may have no ends in $T_s$ other than the two $M$-covered leaves of $T_s$). Then, we cannot apply a (greedy) simple contraction to $\ell_s$ because a simple contraction requires 2 credits. Neither can we contract the semiclosed tree $T_s$ since it is not good.

To get around this difficulty, Nagamochi [10] and Even et al., [9], redefine $M$ to be a maximum matching of the leaf-to-leaf links that are not twin links; we follow this plan in Part II, but in addition to excluding twin links from $M$, we exclude another type of link, the so-called buddy links. Buddy links are defined in Part II; they are associated with a particular configuration that consists of a tree $T_v$ with exactly three leaves and the links incident to $T_v$; each buddy link is associated with a leaf, and this leaf is called a bud. Thus, $M$ is a maximum matching of a subset of the leaf-to-leaf links, where the subset excludes twin links and buddy links.

Redefining $M$ in this way affects the potential function. Recall that the coefficient of an $M$-link is $\frac{3}{2}$ and the coefficient of an $M$-exposed node (a node in $U$) is one, see Lemma 5.1. Observe that each unit "decrease" in $|M|$ "increases" the number of $M$-exposed nodes, $|U|$, by 2, and thus the potential function would "increase" by $\frac{1}{2}$. We cannot allow the potential function to increase (otherwise, Lemma 5.1 could fail since the potential function could exceed $(\frac{3}{2} + \epsilon)y(E)$), hence, we compensate for the increase via the term $\frac{1}{2}\sum_{v \in \mathcal{R}} y(\delta_E(v))$.

Informally speaking, in Part II, a twin link $\ell_s$ contributes $2y(\ell_s)$ to the potential function (rather than $\frac{3}{2}y(\ell_s)$), and we "pay" for the extra $\frac{1}{2}y(\ell_s)$ from the $y$-values of the links incident with the associated stem $s$ of $\ell_s$; similarly, a buddy link $\ell$ contributes $2y(\ell)$ to the potential function, and we "pay" for the extra $\frac{1}{2}y(\ell)$ from the $y$-values of the links incident with the bud $b$ associated with $\ell$.

Although this "fixes" the difficulties caused by excluding twin links and buddy links from $M$, it causes new difficulties. Observe that a stem $s$ may no longer have $\frac{1}{2}y(\delta_E(s))$ credits, hence, the presence of a stem in a semiclosed tree $T_v$ could imply that $T_v$ lacks enough credits to be good. Similarly, a node $w \in \mathcal{R}$ may not have $\frac{1}{2}y(\delta_E(w))$ credits, because it may be incident to a link $bw$ such that $b$ is a bud and we already used $y(bw)$ to "pay" for the buddy link of $b$; hence, the presence of such a node $w$ in a semiclosed tree $T_v$ could imply that $T_v$ lacks enough credits to be good. See Part II for more discussion on these issues.

Three types of "small configurations" have to be handled in Part II, namely, (i) semiclosed trees with three leaves and one stem that are associated with buds and buddy links, (ii) semiclosed trees with four leaves, one stem, and one $M$-link, and (iii) semiclosed trees with four leaves, two stems, and one $M$-link. We use different methods to handle (i)–(iii). As we sketched above, we redefine $M$ and the potential function in order to handle (i), and in addition, we have to handle a cascade of difficulties caused by buddy links; thus, the whole analysis of Part II gets complicated due to the handling of (i). For (ii), either the semiclosed tree is good (so we have sufficient credits to contract it), or it is a so-called deficient 4-leaf tree, and we handle it similarly to the handling of deficient 3-leaf trees in Part I. (In Part II, we show that a semiclosed tree with at least 5 leaves is good, hence, we have only two types of deficient trees, those with three leaves and those with four leaves.) For (iii), either the semiclosed tree is good, or it is a so-called bad 2-stem tree, and we address the latter using a preprocessing step that contracts all bad 2-stem trees; moreover, we have to modify the potential function to ensure that the preprocessing step has sufficient credits for these contractions.

In Part II, the main loop of the algorithm is similar to that of Part I, except that the simple greedy contractions (single-link contractions) of Part I are replaced by (up-to-5) greedy contractions (contractions of a set of links of size $\leq 5$).

In summary, Part II differs significantly from Part I, and some of these differences pertain to the revised potential function, the preprocessing steps, the greedy contractions, the handling of deficient trees, and the handling of buddy links (and side effects).

## 10. Conclusions

In Part I, we proved that the integrality ratio of an SDP relaxation of stemless TAP is $\leq \frac{3}{2} + \epsilon$, where $\epsilon > 0$ can be any small constant. Our analysis is based on the Lasserre tightening of an initial LP. The stemless property is used only in the proof of Theorem 7.3. Part II of the paper goes deeper into the techniques employed in Part I and proves the same result relative to the same SDP relaxation for (general) TAP. The main result of Part II is a polynomial-time algorithm for finding a feasible solution of TAP of size $\leq (\frac{3}{2} + \epsilon)y(E)$, where $y$ is an optimal solution of $\text{Las}_t(LP_0)$, $\epsilon > 0$ is any small constant, and $t = \max\{17, \lceil\frac{1}{2\epsilon}\rceil + 1\}$.

Our methods rely on the Lasserre system and the decomposition theorem. In fact, we do not know how to prove an approximation guarantee of $(\frac{3}{2} + \epsilon)$ for TAP relative to a mathematical programming relaxation *without* using the decomposition theorem (Theorem 4.1). Our critical use of the decomposition theorem is in proving the assertion "if a semiclosed tree $T'_v$ is not good, then $T'_v$ (and its incident links) form a deficient 3-leaf tree," see Theorem 7.3 in Section 7. We proved Theorem 7.3 by applying Theorem 4.1 to decompose a fractional solution $y$ of the Lasserre system into feasible solutions that are integral over a particular set of links $J$ (local integrality). Note that the size of $J$ need not be $O(1)$, and it is possible that $|J| = \Omega(|V|)$. Our proof of Theorem 7.3 makes essential use of the local integrality property on link sets of "unrestricted" size; of course, the key to our approach is to show that $ones(x) \cap J$ has size $O(1)$ for every feasible solution $x$ of our initial LP. (Part II uses the decomposition theorem in a similar way.) Other well known Lift-and-Project systems such as the Lovász-Schrijver system and the Sherali-Adams system are weaker than the Lasserre system, and, to the best of our knowledge, the local integrality property used in our proof of Theorem 7.3 does not hold for $O(1)$ levels of these Lift-and-Project system.

We also use the Lasserre system and the decomposition theorem to derive our potential function, see Section 5, but this use of the decomposition theorem may be "bypassed" because our potential function may be derived using weaker Lift-and-Project systems such as the Lovász-Schrijver system or the Sherali-Adams system. Another way for deriving the potential function is based on formulating a stronger LP relaxation by adding constraints such as the "matching polytope" constraints on the leaf-to-leaf links.

**Acknowledgments.** We thank André Linhares for many discussions. We thank several other colleagues who read preliminary drafts and gave us insightful comments.

Dept. of Comb. & Opt., University of Waterloo, Waterloo, Ontario N2L3G1, Canada.
*E-mail address*: jcheriyan@uwaterloo.ca

*E-mail address*: z9gao@uwaterloo.ca